\documentclass[aps,twocolumn,nofootinbib,showpacs,prd,aps,10pt]{revtex4}
\usepackage[dvips]{graphicx}
\usepackage[english]{babel}
\usepackage[T1]{fontenc}
\usepackage{mathrsfs}
\usepackage[tbtags]{amsmath}
\usepackage{amssymb}
\usepackage{amsxtra}
\usepackage{amsopn}
\usepackage{latexsym}
\usepackage[mathcal]{eucal}
\usepackage{mathtools}

\newcommand{\BE}{\begin{equation}}
\newcommand{\EE}{\end{equation}}
\newcommand{\BA}{\begin{align}}
\newcommand{\EA}{\end{align}}
\newcommand{\Tr}{\mathrm Tr}
\newcommand{\nn}{\nonumber}

\newcommand{\ppp}{ \frac{{\rm d}^4p}{(2\pi)^4}}

\newcommand{\sip}{T\sum_n \int\frac{{\rm d}^3{\bf p}}{(2\pi)^3}}

\begin{document}

\title{Variational study of mass generation and deconfinement 
in Yang-Mills theory}

\author{Giorgio Comitini  and   Fabio Siringo}

\affiliation{Dipartimento di Fisica e Astronomia 
dell'Universit\`a di Catania,\\ 
INFN Sezione di Catania,
Via S.Sofia 64, I-95123 Catania, Italy}

\date{\today}

\begin{abstract}
A very simple variational approach to pure SU($N$) Yang-Mills theory is proposed, 
based on the Gaussian effective potential in a linear covariant gauge. The method
provides an analytical variational argument for mass generation. 
The method can be improved order by order by a perturbative massive expansion around the 
optimal trial vacuum. 
At finite temperature, a weak first-order transition is found (at $T_c\approx 250$ MeV 
for $N=3$) where the mass scale drops discontinuously. 
Above the transition the optimal mass increases
linearly as expected for deconfined bosons. The equation of state is found in good
agreement with the lattice data. 
\end{abstract}


%12.38.Bx       Perturbative calculations
%12.38.Lg	Other nonperturbative calculations (QCD)
%12.38.Aw	General properties of QCD (dynamics, confinement, etc.)
%14.70.Dj	Gluons
%11.15.Tk       Other nonperturbative techniques (gauge field theories)

%12.38.Gc	Lattice QCD calculations (see also 11.15.Ha Lattice gauge theory)
%11.10.Ef       Field Theory: Lagrangian and Hamiltonian approaches
%11.15.-q	Gauge field theories
%12.20.-m       QED
%11.15.Bt       General properties of perturbation theory (gauge theory)
%12.38.-t	Quantum chromodynamics

\maketitle

\maketitle

\section{introduction}
In the last decades the dynamics of QCD has been under intensive theoretical study, aimed at understanding
the properties of matter under the extreme conditions reached by heavy-ion collisions.
Our understanding of the phase diagram has further motivated the study of pure SU($N$) Yang-Mills theory in
the IR and at finite temperature, neglecting quarks as a first approximation. However,
despite the important progresses made, we still miss an analytical description of SU($N$) theory 
from first principles, because of the breaking down of standard perturbation theory below the QCD scale.

The numerical simulation of the theory on a lattice has provided many important insights into the gluon dynamics.
Among them, the dynamical generation of a gluon mass in the dressed 
propagator in the Landau gauge\cite{duarte,cucchieri08,cucchieri08b,bogolubsky,dudal,binosi12,oliveira12,burgio15}
and the occurrence
of a phase transition with the gluons that become deconfined above the critical temperature\cite{lucini,silva,aouane}.
However, since the numerical simulations can only provide data in the Euclidean space, no direct information can be gained
in the Minkowski space where the dynamical properties of the gluon are defined. For instance, no direct proof of confinement
can be obtained on the lattice and even the definition of mass can only be regarded as an energy scale without any clear
dynamical meaning. 

Continuous methods have been developed such as functional 
renormalization group\cite{pawlowski08,pawlowski10,pawlowski10b,pawlowski13}, 
truncation of Dyson-Schwinger equations \cite{epple,alkofer,fischer,papa08,papa10,papa14,papa14b,papa15}
and
Hamiltonian approaches\cite{reinhardt05,reinhardt}. 
They usually require the numerical solution of integral equations and there is no simple way to
extract analytical results from the data.

On the other hand, effective models have been studied analytically, but they are not from first principles and are usually
based on some modified quantization procedure\cite{GZ,dudal08,dudal08b,dudal11}
or different Lagrangians. 
For instance, adding a gluon mass to the Lagrangian is enough for extending the validity
of perturbation theory down to the deep IR, yielding a very good overall picture 
of Yang-Mills theory at one loop\cite{tissier10,tissier11,serreau}. In the context of background field methods the added gluon
mass has provided a good description of the phase diagram at finite temperature, enforcing the idea that most of the non-perturbative
effects can be embedded in the gluon-mass parameter\cite{tissierplb,tissiersu2,tissier15,tissier16}.
While those models are important for understanding the physics of gluons, 
there is a growing interest in the study of analytical approaches to the exact SU($N$) theory.

In this paper, we discuss a very simple variational approach to SU($N$) theory, based on the Gaussian effective potential (GEP)
in a linear covariant gauge. 
We do not modify the original Lagrangian of the theory but optimize the perturbative expansion by a variational argument, yielding
a calculational analytical method that already  provides very important predictions at the lowest orders of the approximation.
Among the main results achieved by the present study we mention: i) a variational argument for mass generation; ii) the
prediction of a first-order deconfinement transition at $T_c\approx 250$ MeV for $N=3$; 
iii) the formal definition of a perturbative expansion around the optimized
vacuum, allowing for an order-by-order improvement of the approximation.

The original approach of Ref.\cite{journey} is here improved and extended to finite temperature, yielding analytical
results up to a one-dimensional numerical integration that is required for the thermal functions.
The perturbative expansion around the vacuum turns out to be the massive expansion developed in Refs.\cite{ptqcd,ptqcd2,analyt,damp}
which was found in excellent agreement with the lattice data\cite{scaling}. Thus, the present study enforces the validity of
that expansion and provides a variational argument for its derivation. Moreover, while by itself the massive expansion cannot
give a genuine proof of mass generation, the variational nature of the GEP can be used as a tool 
for demonstrating that a massless gaussian vacuum of Yang-Mills theory 
is unstable against the vacuum of massive gluons\cite{journey}.

The expansion has been extended to finite temperature in Ref.\cite{damp} allowing for a direct calculation of the gluon
damping rate in the IR and providing a direct proof of confinement. While in that study the zeroth order mass parameter 
was kept fixed, at finite temperature the GEP provides the free energy and allows us to determine the trial mass parameter
variationally, as a function of temperature. The optimal mass scale is found discontinuous at the deconfinement transition, 
leading to an enhancement of the mass decrease that was already found in Ref.\cite{damp},
in agreement with the observed behavior of the Debye mass in lattice simulations\cite{silva}.

The GEP is the energy density of a trial Gaussian vacuum functional that is centered at a given average
value of the field. The width of the functional is given by the mass of the trial free theory
and is determined variationally at each value of the average field, yielding an effective potential
that has been studied by several authors, mainly in the context of spontaneous
symmetry breaking and scalar theories\cite{cornwall74,cornwall82,ibanez,stevenson,stevenson86,stancu2,stancu,reinosa,
gep2,var,light,bubble,su2,LR,HT,superc1,superc2,AF,consoli,stevenson2,stevensonN,branchina}. 
While the GEP is a genuine variational method\cite{stevenson,stevenson86}, several extensions to
higher orders have been proposed\cite{stancu2,stancu,gep2,HT}. 
The idea of an expansion around the optimized vacuum of the GEP 
is not new\cite{tedesco} but has not been developed further.
Expanding around the optimized massive vacuum of the GEP, 
the unconventional massive expansion of Refs.\cite{ptqcd,ptqcd2,analyt} is recovered
in a natural way\cite{journey}. 
Thus, the phenomenological success of the expansion might be due to
the variational choice of a zeroth order vacuum which incorporates most of the non-perturbative effects, 
leaving a residual interaction term that can be treated by perturbation theory.

One of the important merits of the GEP is its paradox of being a pure variational method disguised as a
perturbative calculation,  making use of the standard graphs of perturbation theory. 
Moreover, in the present context, the calculation is highly simplified by the assumption that the average of the gauge field
is zero at the minimum of the potential. 
In other words, we only need the effective potential at its minimum where it is
a function $V(m)$ of the trial mass parameter $m$. However, at variance with 
perturbation theory, the issue of renormalization is less standardized 
in a variational method and the regularization of the diverging integrals 
becomes a central aspect of the calculation.

The paper is organized as follows: in Sec.\ref{sec2} the general formalism
is discussed in the simple case of a scalar theory
where standard well known results are recovered by the method; in Sec.\ref{sec3}
the delicate issue of regularization of the diverging integrals and renormalization
of the GEP is addressed; 
in Sec.\ref{sec4} the GEP for pure SU($N$) Yang-Mills theory is studied
at $T=0$, providing a simple variational argument for mass generation;
in Sec.\ref{sec5} the GEP is extended to finite temperature  and
the phase transition is discussed; a general discussion and a summary of
the results follow in Sec.\ref{sec6}.

\begin{figure}[b] 
\centering
%\vskip 20pt
%\hskip 40pt
\includegraphics[width=0.10\textwidth,angle=-90]{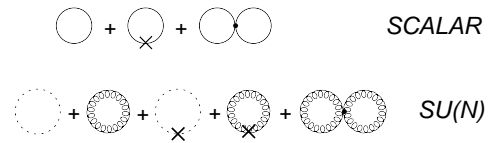}
%\vskip -20pt
\caption{Vacuum graphs contributing to the GEP for the scalar theory (first row) and pure SU($N$) Yang-Mills theory
(second row).}
\label{fig1}
\end{figure}

\section{GEP and mass generation in the scalar theory}\label{sec2}

In order to illustrate the method, in this section
we revise the formalism for the simple case of a self-interacting scalar theory\cite{stevenson}
where the effective potential is well known and is given by three vacuum graphs as shown in Fig.~\ref{fig1}. 
The renormalization scheme will be discussed in the next section.
Most of the arguments developed here are quite general and will be used in the rest of the paper.

Let us consider the Lagrangian 
\BE
{\cal L}=\frac{1}{2} \phi \left(-\partial^2-m_B^2\right)\phi-\frac{\lambda}{4!}\phi^4
\label{Lphi}
\EE
where $m_B$ is a bare mass. We can split the total Lagrangian as ${\cal L}={\cal L}_0+{\cal L}_{int}$
where the trial quadratic part is
\BE
{\cal L}_0=\frac{1}{2} \phi \left(-\partial^2-m^2\right)\phi
\label{Lfree}
\EE
and describes a free scalar particle with a trial mass $m\not= m_B$.
The new interaction follows as
\BE
{\cal L}_{int}=-\frac{\lambda}{4!}\phi^4-\frac{1}{2} \left(m_B^2-m^2\right)\phi^2
\label{Lint}
\EE
so that the total Lagrangian is left unchanged.
If we neglect the interaction, then a free Hamiltonian ${\cal H}_0$ is derived from ${\cal L}_0$ and its
ground state $\vert m\rangle$ satisfies
\BE
{\cal H}_0 \> \vert m\rangle=E_0(m)\> \vert m\rangle
\EE
and depends on the trial mass $m$.
Restoring the interaction ${\cal L}_{int}$, the full Hamiltonian reads ${\cal H}={\cal H}_0+{\cal H}_{int}$ and
by standard perturbation theory, the first-order energy of the ground state reads
\BE
E_1 (m)= E_0(m) + \langle m\vert {\cal H}_{int} \vert m\rangle=\langle m\vert {\cal H} \vert m\rangle
\label{E1}
\EE
and is equivalent to the first-order effective potential $V_1 (m) $ evaluated by perturbation theory in the
covariant formalism with the interaction ${\cal L}_{int}$. Thus, the stationary condition
\BE
\frac{\partial V_1 (m)}{\partial m}=\frac{\partial E_1 (m)}{\partial m}=0
\label{stat}
\EE
gives the best value of $m$ that minimizes the vacuum energy of the ground state $\vert m\rangle$.

While being a pure variational method, the first-order effective potential $V_1(m)=E_1(m)$ can be evaluated 
by the sum of all the vacuum graphs up to first order (the three loop graphs in Fig.~\ref{fig1}). 
The resulting optimized effective potential is the GEP. Usually, the effective potential
is evaluated for any value of the average $\varphi=\langle \phi\rangle$ and the best $m$ also depends on that average.
If the symmetry is not broken, then the minimum of the effective potential is at $\varphi=0$ where
$V_1 (m)$ is a function of the trial mass, to be fixed by the stationary condition Eq.(\ref{stat}).
We assume that the gauge symmetry is not broken in Yang-Mills theory so that $V_1(m)$ at $\varphi=0$ 
is the effective potential 
we are interested in.

The variational nature of the method ensures that the true vacuum energy is smaller than the minimum of $V_1(m)$.
At the minimum, $\vert m\rangle$ provides an approximation for the vacuum and is given by the vacuum of a free
massive scalar particle with mass equal to the optimized mass parameter $m\not=m_B$.
Of course, the optimal state $\vert m\rangle$ is just a first approximation and the actual vacuum is 
much richer. However, we expect that a perturbative expansion around that approximate vacuum would be the
best choice for the Lagrangian ${\cal L}$, prompting towards an expansion with an interaction ${\cal L}_{int}$
and a free part ${\cal L}_0$ that depend on $m$ and can be optimized by a clever choice of the parameter $m$.
Different strategies have been proposed for the optimization, ranging from the stationary condition of the GEP,
Eq.(\ref{stat}), to Stevenson's principle of minimal sensitivity\cite{minimal}. 
A method based on the minimal variance has been recently proposed for QCD and 
other gauge theories\cite{gep2,sigma,sigma2,varqed,varqcd,genself}. 
In all those approaches, the underlying idea is that
an optimal choice of $m$ could minimize the effect of higher orders in the expansion. Since the total Lagrangian
does not depend on $m$, the physical observables are expected to be stationary at the optimal $m$, thus suggesting
the use of stationary conditions for determining the free parameter. As a matter of fact, 
if all graphs were summed up exactly, then the dependence on $m$ would cancel in the final result, so that the 
strength of that dependence measures the weight of the neglected graphs at any order.

Leaving aside the problem of the best choice of $m$, we observe that at $\varphi=0$ the calculation of
the first-order effective potential $V_1(m)$ is quite straightforward and follows from the first-order expansion of
the effective action $\Gamma(\varphi)$
\BE
e^{i\Gamma(\varphi)}=\int_{1PI} {\cal D}_{\phi} e^{i S_0(\phi+\varphi)+i S_{int}(\phi+\varphi)} 
\EE
where the functional integral is the sum of all one-particle irreducible (1PI) graphs and $S=S_0+S_{int}$
is the action. The effective potential then follows as $V(m)=-\Gamma(0)/{\cal V}_4$ where ${\cal V}_4$ is
a total space-time volume. The sum of graphs up to first order gives the first-order effective potential $V_1(m)$
which is the GEP when optimized by Eq.(\ref{stat}).

At finite temperature, the effective potential is replaced by a density of
free energy ${\cal F}(T,m)$ according to
\BE
e^{-\beta \left[{\cal V}_3\>{\cal F}(T,m)\right]}=\int {\cal D}_{\phi} e^{(S_0+S_{int})}
\EE
where the action $S=S_0+S_{int}$ is the integral over imaginary time $\tau$
\BE
S=\int_0^\beta {\rm d}\tau\int {\rm d}^3 x \>{\cal L},
\label{ThS}
\EE
$\beta=1/T$ and ${\cal V}_3$ is a total three-dimensional space volume.
The perturbative expansion of the free energy follows by the same connected graphs contributing to the effective
potential, with loop integrals replaced by a sum over discrete frequencies and a three-dimensional integration.
In the limit $T\to 0$ the effective potential is recovered as $V(m)={\cal F} (0,m)$ and each thermal graph gives
the corresponding vacuum term. Because of the one to one correspondence of the graphs we can easily
switch from the thermal to the vacuum formalism when required.
Moreover, at finite temperature, the GEP maintains its genuine variational nature. In the Hamiltonian formalism, the 
variational argument that follows Eq.(\ref{E1}) can be generalized by Bogolubov's inequality
\BE
{\cal F}\le {\cal F}_0+\frac{1}{{\cal V}_3}
\frac{\Tr\left[{\cal H}_{int}\>\exp (-\beta {\cal H}_0)\right]}
{\Tr\left[\exp (-\beta {\cal H}_0)\right]}={\cal F}_1
\label{BI}
\EE
while in the Lagrangian formalism the same result is found by Jensen-Feynman inequality
\BE
{\cal F}\le {\cal F}_0-\frac{1}{\beta{\cal V}_3}
\frac{\int {\cal D}_{\phi} S_{int}\>e^S_0}{\int {\cal D}_{\phi} \> e^S_0}={\cal F}_1
\label{JF}
\EE
where ${\cal F}_0$ is the free energy obtained by the trial Lagrangian ${\cal L}_0$
while ${\cal F}_1$ is the first order approximation which becomes the GEP when optimized.
The two inequalities tell us that the expansion
must be truncated at first order for a genuine variational approximation.
Here and in the next two sections, when not specified, we will deal with the effective potential and with the renormalization
of the vacuum graphs at zero temperature. The thermal corrections are finite and do not require any
further renormalization.

Since we are interested in the massless Yang-Mills theory, we set $m_B=0$ in the
interaction Eq.(\ref{Lint}) and study a massless scalar theory as a toy model for the problem
of mass generation.
The vertices of the theory can be read from ${\cal L}_{int}$ in Eq.(\ref{Lint}) where we set $m_B=0$ 
and are used in Fig.1 in the vacuum graphs. 
The usual four-point vertex  $-\lambda$ is accompanied by
the counterterm $\delta\Gamma=m^2$ that is denoted by a cross in the graphs. This counterterm must be regarded as part of
the interaction so that the expansion is not loopwise and we find one-loop and two-loop graphs summed together in
the first-order effective potential. That is where the non-perturbative nature of the method emerges since the
expansion in not in powers of $\lambda$ but of the whole interaction ${\cal L}_{int}$.
The zeroth order (massive) propagator $ \Delta_m$ follows from ${\cal L}_0$ 
\BE
\Delta_m (p)=\frac{1}{p^2-m^2}
\EE
and is shown as a straight line in the vacuum graphs.

The tree term is the classical potential and vanishes in the limit $\varphi\to 0$. The first one-loop graph
in Fig.1 gives the standard one-loop effective potential, containing some effects of quantum fluctuations.
It must be added to the second one-loop graph in Fig.1, the crossed graph containing one insertion of the
counterterm. 

It is instructive to see that the exact sum of all one-loop graphs with $n$ insertions of the
counterterm gives the standard vacuum energy of a massless particle. In other words, if we sum all the crossed
one-loop graphs the dependence on $m$ disappears and we are left with the standard one-loop effective potential
of Coleman and Weinberg\cite{WC} $V_{1L}^0=-\Gamma_{1L}^0/{\cal V}_4$ where $\Gamma_{1L}^0$ is the standard
one-loop effective action at $\varphi=0$
\BE
e^{i\Gamma_{1L}^0}=\int {\cal D}_{\phi} e^{i \int \frac{1}{2} \phi (-\partial^2)\phi \>{\rm d}^4x}
\sim\left[{\rm Det}(\Delta_0^{-1})\right]^{\displaystyle{-\frac{1}{2}}}
\label{G1L}
\EE
and $\Delta_0^{-1}=p^2$ is the free-particle propagator of a massless scalar particle.
Up to an additive constant, not depending on $m$, Eq.(\ref{G1L}) can be written as
\BE
V^0_{1L}=\frac{-i}{2{\cal V}_4}\Tr \log(\Delta_m^{-1}+m^2)
\EE
then expanding the log we obtain a {\it massive expansion}
\BE
V^0_{1L}=\frac{-i}{2{\cal V}_4} \Tr\left\{ \log(\Delta_m^{-1})-\sum_{n=1}^\infty
\frac{(-m^2\Delta_m)^n}{n}
\right\}
\label{massexp}
\EE
that is shown pictorially in Fig.2 as a sum of crossed one-loop vacuum graphs. While the sum cannot depend on
$m$, if we truncate the expansion at any finite order we obtain a function of the mass parameter.
As a test of consistency, one can easily check that, once renormalized as described below, the sum of all the crossed 
one-loop vacuum graphs in Fig.2 gives zero exactly.

\begin{figure}[b] 

\vskip 40pt
\hspace*{1cm}\includegraphics[width=0.12\textwidth,angle=-90]{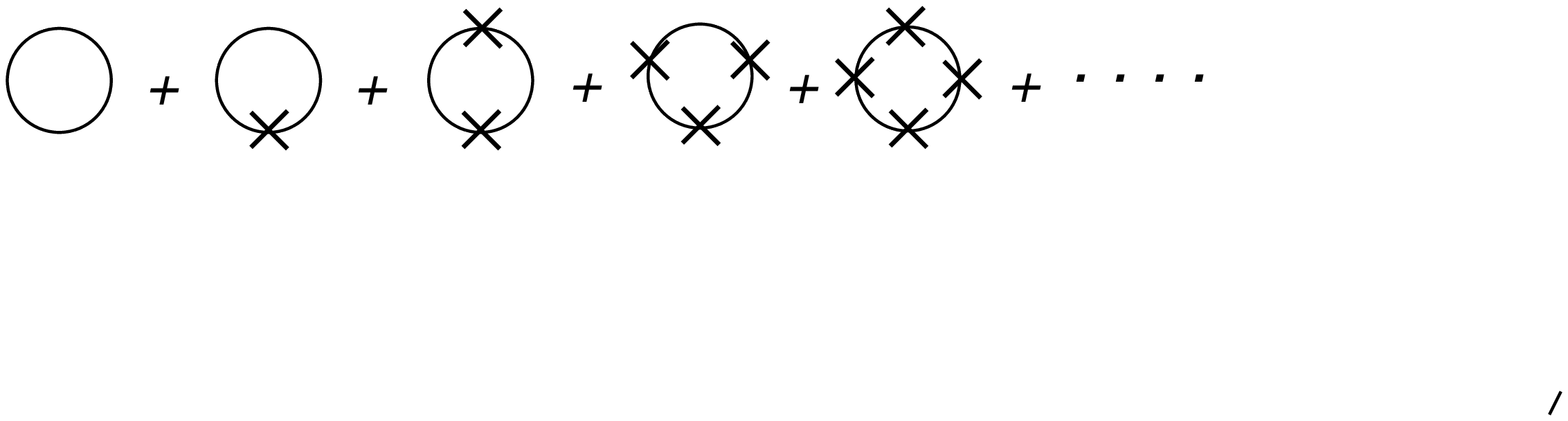}
%\vskip -40pt
\caption{Pictorial display of the right hand side of Eq.(\ref{massexp}).}
\label{fig2}
\end{figure}

The calculation of the GEP requires the sum of only the first two terms of Eq.(\ref{massexp}), 
the two one-loop graphs in Fig.1.  We cannot add
higher-order terms without spoiling the variational method since the average value of the Hamiltonian
in the trial state $\vert m\rangle$ is  $E_1(m)=V_1(m)$, according to Eq.(\ref{E1}).
Using the identity
\BE
\Delta_m=-\frac{\partial}{\partial m^2} \log(\Delta_m^{-1})
\label{ide}
\EE
the sum of one-loop graphs in Fig.1 can be written as
\BE
V_{1L} (m)=\left( 1-m^2\frac{\partial}{\partial m^2}\right) K(m)=K(m)-\frac{1}{2} m^2 J(m)
\label{V1L}
\EE
where $K(m)$ and $J(m)$ are defined as
\begin{align}
K(m)&=\frac{-i}{2{\cal V}_4}\Tr \log(\Delta_m^{-1})\nn\\
J(m)&=\frac{i}{{\cal V}_4}\Tr \Delta_m
\label{KJ}
\end{align}
and because of Eq.(\ref{ide}), satisfy the identity
\BE
\frac{\partial K (m)}{\partial m^2}=\frac{1}{2} J(m).
\label{dI1}
\EE
At $T=0$ they can be written as explicit
diverging integrals
\begin{align}
K(m)&=\frac{1}{2i}\int \ppp \log(-p^2+m^2)\nn\\
J(m)&=-i\int \ppp \frac{1}{-p^2+m^2}
\label{I}
\end{align}
to be regularized in some renormalization scheme.
At finite temperature Eq.(\ref{dI1}) still holds, but
the integrals acquire a finite additive thermal part.

We recognize $K (m)$ as the standard one-loop effective potential of  Coleman and Weinberg for a
massive scalar particle in the limit $\varphi\to 0$. This term contains the quantum fluctuations at
one-loop. The second term in Eq.(\ref{V1L}) is a correction coming from the counterterm and arises because the exact
Lagrangian was massless. 

The calculation of the GEP also requires the two-loop graph in Fig.1 which
is first-order in $\lambda$. It can be recovered from the crossed one-loop graph
by just substituting the vertex $-m^2$  
with the seagull one-loop self energy $ \Sigma_{1L}$ that reads\cite{gep2}
\BE
\Sigma_{1L}=\frac{\lambda}{2} J (m)
\label{sea}
\EE
and adding a $1/2$ symmetry factor. The resulting two-loop term is
\BE
V_{2L} (m)=\frac{\lambda}{8} [J (m)]^2.
\label{V2L}
\EE
The GEP follows as the sum $V_{1L}+V_{2L}$
\BE
V_{G}(m)=K(m)-\frac{1}{2}m^2 J(m) +\frac{\lambda}{8} [J (m)]^2.
\label{GEP}
\EE
At this stage we have just recovered the GEP in the limit $\varphi\to 0$ and Eq.(\ref{GEP}) agrees with
the well known GEP in that limit\cite{stevenson,stancu2,stancu,reinosa,gep2}. 

More precisely, $V_{G}$ is the GEP when $m$ is optimized by the stationary condition Eq.(\ref{stat})
that reads
\BE
\frac{\partial V_{G} (m)}{\partial m^2}=\frac{1}{2}\left(\frac{\partial J (m)}{\partial m^2}\right)
\left[\frac{\lambda J(m)}{2}-m^2\right]=0
\label{factors}
\EE
yielding the usual gap equation of the GEP
\BE
m^2=\frac{\lambda J(m)}{2}.
\label{gap}
\EE
From a mere formal point of view, if Eq.(\ref{gap}) has a non-zero solution, the GEP predicts the existence of a mass
for the massless scalar theory.
That is of special interest because for $m_B=0$ the Lagrangian in Eq.(\ref{Lphi}) has no energy scale, just
like Yang-Mills theory and QCD in the chiral limit. Thus, it can be regarded as a toy model for the more
general problem of mass generation and chiral symmetry breaking.

\section{Renormalization of the GEP}\label{sec3}

The scalar theory has been studied by many authors in the past, using
different regulators, ranging from the insertion of a cut-off to dimensional regularization and,
of course, to lattice regularization. The resulting physical theories are not always equivalent and the problem of
triviality is still not totally solved. The issue is quite subtle and has to do with the physical meaning that we give
to the theory in a four dimensional space. The regularization of the GEP has also been addressed 
by many methods\cite{cornwall74,stevenson,consoli,stevenson2,stevensonN,branchina,reinosa}.
The most intuitive way of regularizing the integrals is by inserting a large but finite cutoff $\Lambda$ which
provides the physical units of the theory, as in lattice calculations where the finite lattice spacing $a$
cuts the energies larger than $\Lambda\sim 1/a$. In the Euclidean space, the integral $J$ reads
\BE
J(m)=\int_0^{\Lambda^2}\frac{p^2{\rm d}p^2}{16\pi^2}  
\left[\frac{1}{p^2+m^2}\right]\> >\> 0
\label{JUV}
\EE
and is a finite positive-definite function of the mass parameter. 
The gap equation, Eq.(\ref{gap}), has a well defined solution at $m^2=m^2_0=c_\lambda \lambda \Lambda^2/(32\pi^2)$
where $c_\lambda$ is a coefficient of order unity, with  $0<c_\lambda<1$ and $c_\lambda\approx 1$ in the limit $\lambda\to 0$.
Since the derivative 
\BE
\frac{\partial J(m)}{\partial m^2} < 0
\label{dJUV}
\EE
is negative for any value of $m^2$, the derivative of the effective potential
in Eq.(\ref{factors}) changes sign at $m=m_0$ and becomes positive for $m>m_0$. Thus, the GEP has an absolute minimum
at $m_0$ and the simple cut-off regularization predicts a mass. 
The existence of a minimum at $m=m_0>0$ makes sense when compared with the data of lattice simulations that
predict the existence of a finite mass in the limit $m^2_B\to 0^+$ of the unbroken-symmetry theory\cite{huang87}.
However, that mass is not a dynamical mass and arises from the quadratic divergence of $J$ because no special
symmetry protects the theory. That is not a desirable feature in a toy model for Yang-Mills theory since
BRST invariance, which is not broken on the lattice, forbids the appearance of diverging mass terms.
In that context, dimensional regularization is the first choice since it leaves BRST unbroken and is
the simplest and usual way to cancel the quadratic divergence. 

Having set $d=4-\epsilon$, in the limit $\epsilon\to 0$
the integral $J$ is redefined as $J\mu^\epsilon$ where $\mu$ is an arbitrary scale of the order of $m$ and
expanding in powers of $\epsilon$
\BE
J(m)=-\frac{m^2}{16\pi^2}\left[\frac{2}{\epsilon} + \log \frac{\bar\mu^2}{m^2}+1+{\cal O}(\epsilon)\right]
\EE
where $\bar \mu=(2\sqrt{\pi}\mu)\exp(-\gamma/2)$. 
Integrating Eq.(\ref{dI1})
and neglecting an integration constant (that does not depend on $m$)
\BE
K(m)=-\frac{m^4}{64\pi^2}\left[\frac{2}{\epsilon} + \log \frac{\bar\mu^2}{m^2}+\frac{3}{2}+{\cal O}(\epsilon)\right].
\EE

In the usual approach of Coleman and Weinberg\cite{WC}, the divergences are absorbed by the (infinite) 
integration constants that are traded as finite and physical renormalized parameters. 
Following that approach, we could hide the poles in the definition of an energy scale $\Lambda_\epsilon$ such that 
\BE
\log \Lambda_\epsilon^2= \log \bar \mu^2+\frac{2}{\epsilon}+1
\label{Lambda}
\EE
and write the integrals $K$, $J$ as simply as
\begin{align}
J(m)&=\frac{m^2}{16\pi^2}\log \frac{m^2}{\Lambda_\epsilon^2}\nn\\
K(m)&=\frac{m^4}{64\pi^2}\left[\log \frac{m^2}{\Lambda_\epsilon^2}-\frac{1}{2}\right].
\label{IDR}
\end{align}
If $\Lambda_\epsilon$ were traded as a finite unknown energy scale, then the regularized expressions of $J$ and $K$
would be finite.  

Let us investigate the limits of Eq.(\ref{IDR}) when
the definition of $\Lambda_\epsilon$, Eq.(\ref{Lambda}), is taken literally, in the attempt to give it a physical meaning.
While $\epsilon$ might even be a complex variable and the physical meaning of the poles is quite obscure in general,
Eq.(\ref{Lambda}) only makes sense if we assume that $\epsilon$ is real, at least. Moreover, the expansion
can only be trusted if $\vert \epsilon \log  (\bar \mu^2/m^2)\vert\ll 1$ which is equivalent to say that
\BE
\log\frac{\Lambda_\epsilon^2}{m^2}\approx \frac{2}{\epsilon}\to \pm\infty
\EE
yielding $m\ll \Lambda_\epsilon$ if $\epsilon>0$ and $m\gg \Lambda_\epsilon$ if $\epsilon<0$.
Thus, if we literally assume to work in a $(4\mp\vert\epsilon\vert)$-dimensional spacetime, Eq.(\ref{IDR}) holds
asymptotically for a very small or a very large mass compared to $\Lambda_\epsilon$. The energy scale $\Lambda_\epsilon$ can be regarded as a
very large UV cutoff or a very small IR cutoff, according to the sign of $\epsilon$.
In both cases, we must face the non-intuitive result that the regularized $J$ and its derivative change sign according
to the value of $m$: for $m\ll \Lambda_\epsilon$ the integral $J$ is negative while for $m\gg\Lambda_\epsilon$ the derivative of $J$ becomes
positive, which is at odds with the intuitive result obtained by a simple cutoff in Eqs.(\ref{JUV}),(\ref{dJUV}).
Actually, we must recognize that dimensional regularization is not neutral but its way to make sense of
divergences is part of the physical interpretation of a field theory, with scaleless integrals that vanish exactly
and a less marked difference between UV and IR divergences.
Moreover, the use of dimensional regularization is
controversial in the scalar theory and different physical theories seem to arise when the limit $d\to 4$ is taken
from above ($d>4$) or below ($d<4$), as first pointed out by Stevenson\cite{stevenson2} in 1987. While it is
still not obvious if any of them describes the lattice-regulated scalar theory, they could be very relevant for our toy
model of Yang-Mills theory. After reviewing them briefly, we will show how a dimensional regularization scheme can be set up for the variational
effective potential of Yang-Mills theory.

\subsection{The autonomous theory ($d<4$)}

The autonomous renormalization of scalar theory\cite{stevenson,consoli} can be easily recovered by dimensional
regularization for $d<4$~\cite{stevenson2,branchina}. It shows spontaneous symmetry breaking and asymptotic freedom but
cannot be connected, perturbatively, to the usual low energy phenomenology that emerges 
by perturbation theory and $1/N$ expansion\cite{stevenson2,stevensonN}.

The search for a minimum of the GEP yields the coupled equations\cite{consoli,branchina}
\begin{align}
m_0^2&=\frac{1}{3}\lambda \varphi_0^2\nn\\
m_0^2&=-\frac{\lambda}{2} J(m_0)
\label{SSB}
\end{align}
where $\varphi_0$ is the optimal average value of the scalar field that would eventually break the symmetry if a solution exists.
In that case, the other stationary point at $\varphi=0$ is a maximum where Eq.(\ref{gap}) holds. If the symmetry is broken Eq.(\ref{gap})
is replaced by the second of Eqs.(\ref{SSB}), which has the opposite sign and has a physical solution if $\epsilon\to 0^+$ ($d<4$).
In fact, using the first of Eqs.(\ref{IDR}), the new gap equation reads
\BE
\frac{1}{\alpha}=\log\frac{\Lambda_\epsilon}{m_0}
\label{alphab}
\EE
where $\alpha=\lambda/(16\pi^2)$ is a bare effective coupling and $\Lambda_\epsilon\to\infty$ in the limit $\epsilon\to 0^+$
so that $\alpha\to 0^+$ is positive. The solution $m_0$ of the gap equation can be regarded as a physical scale which breaks the
symmetry according to the first of Eqs.(\ref{SSB}). Assuming that $m_0$ takes some fixed phenomenological value, the large scale $\Lambda_\epsilon$
can be eliminated as
\BE
\Lambda_\epsilon=m_0\> e^{1/\alpha}
\label{gap+}
\EE
so that the theory shows asymptotic freedom. Inserting the explicit expressions of $J$ and $K$ in the effective potential, the GEP at its
minimum is\cite{consoli,branchina}
\BE
V_G=-\frac{m_0^4}{128 \pi^2} <0
\label{VGA}
\EE
and $\Lambda_\epsilon$ can be sent to infinity ($\epsilon\to 0^+$) yielding a finite energy density, spontaneous symmetry breaking 
and a finite physical mass $m_0$.

\begin{figure}[b] 
\centering
\includegraphics[width=0.3\textwidth,angle=-90]{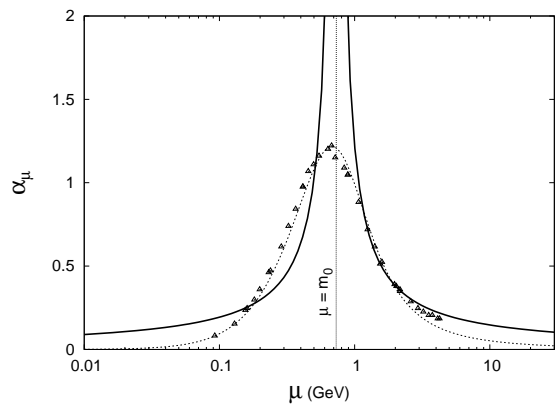}
\caption{The running coupling $\alpha_\mu$ of Eq.(\ref{RGUV3}) is shown for $m_0=0.73$ GeV (solid line), together
with the lattice data of Ref.\cite{bogolubsky} for the strong coupling $\alpha_s$ of Yang-Mills theory in the Taylor scheme.
The exponent $\nu$ is arbitrarily fixed by matching the data at $\mu=2$ GeV. The dotted line is the analytical result
of Ref.\cite{ptqcd2}, obtained by a one-loop expansion around the Gaussian massive vacuum at $m=m_0=0.73$ GeV.} 
\label{fig3}
\end{figure}

At variance with perturbation theory, in principle, the variational method does not require the use of a renormalized coupling.
However, it is useful to parametrize the gap equation in terms of a finite running coupling $\alpha_\mu$ which can be defined
according to\cite{branchina,reinosa}
\BE
\frac{1}{\alpha}=\frac{1}{\alpha_\mu} +\log\frac{\Lambda_\epsilon}{\mu}>0
\label{alphaUV}
\EE
where $\mu$ is any finite scale. The gap equation, Eq.(\ref{alphab}), is written as a finite renormalized gap equation
\BE
\frac{1}{\alpha_\mu}=\log\frac{\mu}{m_0}
\label{RGUV}
\EE
where $m_0$ is assumed to be the physical RG invariant mass. As a toy model of Yang-Mills theory, we assume that $\alpha_\mu>0$,
so that $\mu$ must be larger than $m_0$ and the running of $\alpha_\mu$ takes place in the UV sector, limited from below by 
the Landau pole at $\mu=m_0$.
The beta function is negative and the running coupling shows asymptotic freedom. 
A plot of the coupling $\alpha_\mu$ is shown (up to a factor) as a solid line on the right side of Fig.~3.
The breaking of symmetry and the existence of a mass scale seem
to reverse the usual trivial behavior of the scalar theory. The autonomous behavior is separated from the usual weak coulpling limit
which is observed below the Landau pole.
However, we must mention that Eq.(\ref{RGUV}) is just a possible reparametrization of Eq.(\ref{alphab}); it is not necessary, since the
effective potential is anyway RG invariant at its minimum; and besides, the parametrization is not unique. It has some features that make it a good candidate as a physical
renormalized coupling at the scale $\mu$: in fact, reversing Eq.(\ref{alphaUV}) it can be written in the perturbative weak coupling
limit as $\alpha_\mu=\alpha [1+{\cal O}(\alpha)]$ and $\alpha_\mu\to \alpha$ in the UV limit $\mu\to \Lambda_\epsilon$. But, 
it is not obvious how $\alpha_\mu$ is related to the four-point function at the scale $\mu$. Moreover, the parametrization is not unique:
the existence of a RG invariant energy scale $m_0$ allows us to define a generic scale 
\BE
\Lambda^\prime_\epsilon=m_0 \left(\frac{\Lambda_\epsilon}{m_0}\right)^\nu
\EE
and a different running coupling $\alpha^\prime_\mu$ according to
\BE
\frac{1}{\alpha}=\frac{1}{\alpha^\prime_\mu} +\frac{1}{\nu}\log\frac{\Lambda^\prime_\epsilon}{\mu}
\label{alphaUV2}
\EE
yielding by Eq.(\ref{alphab}) the finite equation
\BE
\frac{1}{\alpha^\prime_\mu}=\frac{1}{\nu}\>\log\frac{\mu}{m_0}.
\label{RGUV2}
\EE
Thus, the coefficient of the beta function is somehow arbitrary and we do not expect that any serious prediction can be made without
an explicit calculation of the four-point function. Quite interesting, the exponent $\nu$ can be taken negative, inverting the sign
of the beta function. However, assuming that $\alpha^\prime_\mu>0$, we obtain $\mu< m_0$ if $\nu<0$. The negative beta function
would be defined below the Landau pole, and the new parametrization would describe the IR sector of the theory showing the same behavior
that is predicted by perturbation theory and $1/N$ expansion: an increasing running coupling and triviality.
For a negative $\nu$, a plot of $\alpha^\prime_\mu$ is shown as a solid line on the left side of Fig.~3.
We observe that if $\nu<0$ then $\Lambda^\prime_\epsilon\to 0$ in the limit $\epsilon\to 0^+$ when $\Lambda_\epsilon\to\infty$. 
Let us consider the special case $\nu=-1$ and call
$\delta_\epsilon=\Lambda^\prime_\epsilon$ in order to make clear that it is an infinitesimal IR scale, $\delta_\epsilon\to 0$.
Eq.(\ref{alphab}) can be written as
\BE
\frac{1}{\alpha}=\log\frac{m_0}{\delta_\epsilon}
\label{alphab2}
\EE
which has the same identical content as before, but in terms of the IR vanishing scale $\delta_\epsilon =m_0\exp(-1/\alpha)$.
Thus the same theory now looks trivial. It is important to see that different parametrizations for $\nu=\pm 1$, predicting
opposite beta functions, refer to different ranges of $\mu$, separated by the Landau pole. Thus the respective weak coupling limits
cannot be connected by perturbation theory, yielding a double-valued beta which is legitimate when the running coupling is not a monotone function.
In fact, joining together the outcome of Eq.(\ref{RGUV2}) for $\pm \nu$ we obtain 
\BE
\frac{1}{\alpha^\prime_\mu}=\left\vert\frac{1}{\nu}\log\frac{\mu}{m_0}\right\vert
\label{RGUV3}
\EE
which holds for any $\mu\not=m_0$, as shown in Fig.~3 where $\vert \nu\vert $ is arbitrarily chosen to match the strong coupling $\alpha_s$ at
$\mu=2$ GeV.

\subsection{The precarious theory ($d>4$)}

Despite its name, the precarious renormalization of scalar theory\cite{stevenson} predicts the same phenomenology
of perturbation theory and $1/N$ expansion\cite{stevensonN}. Its handling by a cut-off is problematic since it
seems to be unstable until the cut-off is sent to infinite. It emerges in a natural and straightforward way by dimensional
regularization in $d>4$, as first shown by Stevenson\cite{stevenson2}. 

In the limit $\epsilon\to 0^-$, the energy scale $\Lambda_\epsilon$ goes to zero according to Eq.(\ref{Lambda}).
Let us call it $\delta_\epsilon$ in order to make clear that $\delta_\epsilon=\Lambda_\epsilon\to 0$.
In the same limit, the coupled equations for the minimum of the GEP, Eqs.(\ref{SSB}), have no solution because the bare coupling
$\alpha$ would become negative in Eq.(\ref{alphab}). There is no spontaneous symmetry breaking 
and the minimum of the effective potential is at $\varphi=0$.
At that point, having ruled out the breaking of symmetry, Eq.(\ref{gap}) holds and can be written as
\BE
\frac{1}{\alpha}=\log\frac{m_0}{\delta_\epsilon}
\label{alphab3}
\EE
which has the opposite sign of Eq.(\ref{alphab}).
In the limit $\delta_\epsilon\to 0$ the bare coupling $\alpha$ is positive and an acceptable solution $m_0$ exists. As before,
we assume that $m_0$ is a RG invariant physical mass which is generated dynamically in the massless theory. Thus, the small energy
scale $\delta_\epsilon$ can be eliminated as $\delta_\epsilon =m_0\exp(-1/\alpha)$ in the effective potential.
We observe that Eq.(\ref{alphab3}) is identical to Eq.(\ref{alphab2}), and the theory appears as trivial.

\begin{figure}[b] 
\centering
\includegraphics[width=0.3\textwidth,angle=-90]{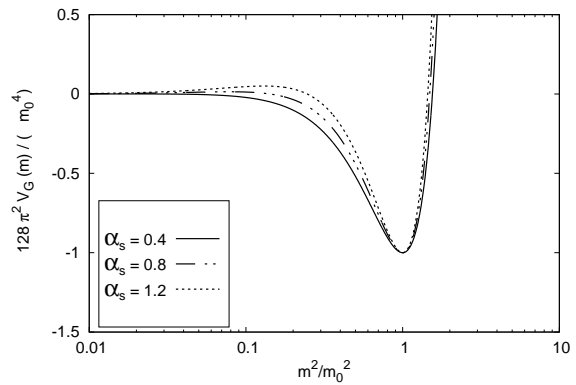}
\caption{The renormalized GEP  of Eq.(\ref{GEPsc}) is shown in units of $m_0$ for different values
of the strong coupling $\alpha_s$, having set $\alpha=\frac{9N}{8\pi}\alpha_s$.}  
\label{fig4}
\end{figure}

At its minimum $\varphi=0$, the effective potential is given by Eq.(\ref{GEP}) and inserting the regularized expressions of
the integrals $J$, $K$, as given by Eqs.(\ref{IDR}) with $\Lambda_\epsilon=\delta_\epsilon\to 0$, we can write it as
\BE
V_{G}(m)=\frac{m^4}{128\pi^2} \left[\alpha \left(\log \frac{m^2}{\delta_\epsilon^2}\right)^2-2\log\frac{m^2}{\delta_\epsilon^2} -1\right]
\label{U}
\EE
which obviously makes sense only if $m\gg \delta_\epsilon$. Eliminating $\delta_\epsilon$ by Eq.(\ref{alphab3}) the renormalized
GEP reads
\BE
V_{G}(m)=\frac{m^4}{128\pi^2} \left[\alpha \left(\log \frac{m^2}{m_0^2}\right)^2+2\log\frac{m^2}{m_0^2} -1\right]
\label{GEPsc}
\EE 
and is shown in Fig.~4. The only physical point is the absolute  minimum at $m^2=m_0^2$ where the effective potential does not
depend on the bare coupling $\alpha$ and takes the value
\BE
V_{G}(m_0)=-\frac{m_0^4}{128\pi^2}<0.
\label{VG0}
\EE
Then we can safely send $\epsilon\to 0$. We obtain the same identical vacuum energy that was found in Eq.(\ref{VGA}) by the autonomous 
renormalization in $d<4$, but here the mass $m_0$ is generated without any symmetry breaking.

We observe that the stationary point $m_0$ is the physical mass that emerges as the pole of the self-consistent propagator.
Actually, up to first order, the self-energy is the sum of the tree-level counterterm $-m^2$ and the seagull 
graph $\Sigma_{1L}$ in Eq.(\ref{sea}), so that the self-consistency condition $m=m_0$ is equivalent to the vanishing of
the first-order self energy\cite{gep2}
\BE
\Sigma_{1}=-m^2+\frac{\lambda}{2} J (m)=0
\EE
which is just the stationary condition Eq.(\ref{gap}) satisfied by $m_0$.

As discussed for $d<4$, we do not need to introduce any running coupling in the variational calculation, because
the effective potential is finite in units of $m_0$. However, it might be useful to reparametrize the gap equation
by a finite running coupling $\alpha_\mu$ which can be defined as before\cite{stevenson2}
\BE
\frac{1}{\alpha}=\frac{1}{\alpha_\mu} +\log\frac{\mu}{\delta_\epsilon}>0
\label{alphaIR}
\EE
where $\mu$ is an arbitrary energy scale. The gap equation, Eq.(\ref{alphab3}), is then written as a finite renormalized gap equation
\BE
\frac{1}{\alpha_\mu}=\log\frac{m_0}{\mu}
\label{RGIR}
\EE
where $m_0$ is the physical RG invariant mass.
Since we assume that $\alpha_\mu>0$, here $\mu$ must be smaller than $m_0$ and the running of $\alpha_\mu$ takes place in the IR sector, 
below the Landau pole at $\mu=m_0$. While we could deduce, naively, that the theory is trivial and the beta function is positive, 
again we must recognize that the parametrization is not unique and the running of $\alpha_\mu$ is limited in the IR sector.
In fact, Eq.(\ref{RGIR}) is identical to Eq.(\ref{RGUV2}) for $\nu=-1$ and the present theory gives the same running 
predicted by the autonomous theory in the IR sector.
Again, the existence of the RG invariant mass $m_0$ allows us to define a new energy scale 
\BE
\Lambda^\prime_\epsilon=m_0 \left(\frac{\delta_\epsilon}{m_0}\right)^\nu
\EE
and a different running coupling $\alpha^\prime_\mu$ according to
\BE
\frac{1}{\alpha}=\frac{1}{\alpha^\prime_\mu} +\frac{1}{\nu}\log\frac{\mu}{\Lambda^\prime_\epsilon}
\label{alphaIR2}
\EE
yielding by Eq.(\ref{alphab3}) the finite equation
\BE
\frac{1}{\alpha^\prime_\mu}=\frac{1}{\nu}\>\log\frac{m_0}{\mu}.
\label{RGIR2}
\EE
Joining together the outcome of Eq.(\ref{RGIR2}) for $\pm \nu$ we obtain the same identical result of Eq.(\ref{RGUV3}) which holds
for any $\mu\not=m_0$ and is shown as a solid line in Fig.~3. 
We conclude that, up to an unknown factor $\nu$, the beta function might have the same behavior in both 
renormalization schemes.

\subsection{A toy model for Yang-Mills theory}

When regularized dimensionally, two different renormalized theories seem to emerge in the limit $d\to 4$.
However, for many aspects, the two renormalized theories appear as two sides of the same coin.
Both theories share a dynamical mass generation, the same vacuum energy density, a Landau pole at $\mu=m_0$
and can be parametrized by the same running coupling which is not monotone, showing asymptotic freedom in the UV and
a trivial Gaussian fixed point in the IR. 

In both cases the Landau pole that emerges in the reparametrization 
has no effect on the effective potential which is RG invariant and is valid at any energy scale. Actually,
at variance with perturbation theory, the variational method does not even require the use of a running coupling.
However, the existence of the pole says that the two weak-coupling limits cannot be connected by perturbation
theory which must break down at the scale $\mu\approx m_0$. In fact, by general arguments, perturbation theory predicts 
that the beta function must be unique at the lowest orders of approximation and cannot depend on the special
regularization scheme. But, if the running coupling is not a monotone function, a double valued beta function is
found, taking different (opposite) values in different sectors that cannot be connected by perturbation theory.
That scenario is only compatible with the existence of a RG invariant phenomenological energy scale where
perturbation theory breaks down.

If we look at the strong coupling $\alpha_s$ of Yang-Mills theory in the Taylor scheme, a non-monotonic behavior
is found in the Landau gauge on the lattice\cite{bogolubsky}, assuming that the ghost-gluon vertex is
regular and a running coupling can be defined from the product of the dressing functions of two-point correlators.
Some lattice data of Ref.\cite{bogolubsky} are shown in Fig.~3 together with the analytical prediction of Ref.\cite{ptqcd2},
obtained by a one-loop massive expansion around the zeroth-order Gaussian propagator $(-p^2+m_0^2)^{-1}$ with $m_0=0.73$ GeV.

The energy $\mu\approx 0.7$ GeV, where the coupling reaches its maximum, is the
phenomenological scale where perturbation theory breaks down. Somehow, the running coupling $\alpha_\mu$ of Eq.(\ref{RGUV3})
can be seen as a zeroth-order Gaussian approximation for the strong coupling $\alpha_s(\mu)$ of  Yang-Mills theory.
Actually, that is no coincidence since a gauge invariant effective potential will be derived in the next section for Yang-Mills
theory, which is exactly the same GEP of Eq.(\ref{GEPsc}) and Fig.~4, apart from a normalization factor and
the precise definition of the effective coupling $\alpha$. Thus, irrespective of the agreement with the lattice-regulated
scalar theory, the dimensional-regulated GEP of scalar theory is a useful toy model for pure Yang-Mills theory.

The two scalar theories only differ because of the breaking of symmetry which appears for $d<4$;
while, for $d>4$, a dynamical mass generation occurs without any symmetry breaking. Since gauge symmetry is not broken in Yang-Mills
theory, we expect that the correct phenomenology can only be reproduced if we adopt the second scheme and regularize the theory
keeping $d>4$.

\section{GEP and mass generation in SU($N$) Theory}\label{sec4}

The Lagrangian of pure SU($N$) Yang-Mills theory can be written as
\BE
{\cal L}={\cal L}_{YM}+{\cal L}_{fix}+{\cal L}_{FP}
\label{LYM}
\EE
where ${\cal L}_{YM}$ is the Yang-Mills term
\BE
{\cal L}_{YM}=-\frac{1}{2} \Tr\left(  \hat F_{\mu\nu}\hat F^{\mu\nu}\right)
\EE
${\cal L}_{fix}$ is a gauge fixing term and ${\cal L}_{FP}$ is the ghost Lagrangian
arising from the Faddev-Popov determinant.
In terms of the gauge fields, the tensor operator $\hat F_{\mu\nu}$ is 
\BE
\hat F_{\mu\nu}=\partial_\mu \hat A_\nu-\partial_\nu \hat A_\mu
-i g \left[\hat A_\mu, \hat A_\nu\right]
\EE
where
\BE
\hat A^\mu=\sum_{a} \hat T_a A_a^\mu
\EE
and the generators of SU($N$) satisfy the algebra
\BE
\left[ \hat T_a, \hat T_b\right]= i f_{abc} \hat T_c
\EE
with the structure constants normalized according to
\BE
f_{abc} f_{dbc}= N\delta_{ad}.
\label{ff}
\EE
If a generic linear covariant gauge-fixing term is chosen
\BE
{\cal L}_{fix}=-\frac{1}{\xi} \Tr\left[(\partial_\mu \hat A^\mu)(\partial_\nu \hat A^\nu)\right],
\EE
where $\xi>0$ is an arbitrary positive number,
the total action can be written as ${\cal S}_{tot}={\cal S}_0+{\cal S}_I$ where the free-particle term is
\begin{align}
{\cal S}_0&=\frac{1}{2}\int A_{a\mu}(x)\delta_{ab}\> {\Delta_0^{-1}\>}^{\mu\nu}(x,y)\> A_{b\nu}(y) 
{\rm d}^dx{\rm d}^dy \nn \\
&+\int \omega^\star_a(x) \delta_{ab}\>{{\cal G}_0^{-1}}(x,y) \>\omega_b (y) {\rm d}^dx{\rm d}^dy
\label{S0}
\end{align}
and the interaction is
\BE
{\cal S}_I=\int{\rm d}^dx \left[ {\cal L}_{3g} +   {\cal L}_{4g}+{\cal L}_{gh} \right]
\label{SI}
\EE
with the usual local interaction terms that read
\begin{align}
{\cal L}_{3g}&=-g  f_{abc} (\partial_\mu A_{a\nu}) A_b^\mu A_c^\nu\nn\\
{\cal L}_{4g}&=-\frac{1}{4}g^2 f_{abc} f_{ade} A_{b\mu} A_{c\nu} A_d^\mu A_e^\nu\nn\\
{\cal L}_{gh}&=-g f_{abc} (\partial_\mu \omega^\star_a)\omega_b A_c^\mu.
\label{LintYM}
\end{align}
In Eq.(\ref{S0}), $\Delta_0$ and ${\cal G}_0$ are the standard free-particle propagators for
gluons and ghosts and their Fourier transforms are
\begin{align}
{\Delta_0}^{\mu\nu} (p)&=\Delta_0(p)\left[t^{\mu\nu}(p)  
+\xi \ell^{\mu\nu}(p) \right]\nn\\
\Delta_0(p)&=\frac{1}{-p^2}, \qquad {{\cal G}_0} (p)=\frac{1}{p^2}.
\label{D0}
\end{align}
Here the transverse and longitudinal projectors are defined as
\BE
t_{\mu\nu} (p)=g_{\mu\nu}  - \frac{p_\mu p_\nu}{p^2};\quad
\ell_{\mu\nu} (p)=\frac{p_\mu p_\nu}{p^2}
\label{tl}
\EE
where $g_{\mu\nu}$ is the metric tensor. 

As discussed in Refs.\cite{ptqcd2,analyt}, an unconventional massive expansion can be introduced
by adding and subtracting mass terms $\delta {\cal S}_i$ in the total action, just like we did for the
scalar theory in Eqs.(\ref{Lfree}),(\ref{Lint}).
The method can be generalized by redefining the free and interacting parts of the action
\begin{align}
{\cal S}_0&\rightarrow {\cal S}_0-\sum_i\delta {\cal S}_i\nn\\
{\cal S}_I&\rightarrow {\cal S}_I+\sum_i\delta {\cal S}_i.
\label{shift}
\end{align}
For the gluon we can take
\BE
\delta {\cal S}_g= \frac{1}{2}\int A_{a\mu}(x)\>\delta_{ab}\> \delta\Gamma^{\mu\nu}(x,y)\>
A_{b\nu}(y) {\rm d}^dx{\rm d}^dy 
\label{dS}
\EE
where the vertex function $\delta\Gamma^{\mu\nu}$ is given by
a shift of the inverse propagator
\BE
\delta \Gamma^{\mu\nu}(x,y)=
\left[{\Delta_0^{-1}}^{\mu\nu}(x,y)- {\Delta_m^{-1}}^{\mu\nu}(x,y)\right]
\label{dGg}
\EE
and ${\Delta_m}^{\mu\nu}$ is the massive free-particle propagator
\begin{align}
{\Delta_m^{-1}}^{\mu\nu} (p)&=\Delta_m^T(p)^{-1} t^{\mu\nu}(p)  
+\Delta_m^L(p)^{-1} \ell^{\mu\nu}(p)\nn\\
\Delta_m^T(p)&=\frac{1}{-p^2+m^2}, \qquad \Delta_m^L(p)=\frac{\xi}{-p^2+m_L^2}
\label{Deltam}
\end{align}
As a general variational ansatz, the two masses $m$ and $m_L$ can be different.

In principle, we would also have the freedom to insert a mass shift $\delta {\cal S}_{gh}$ for the ghost 
\BE
\delta {\cal S}_{gh}= \int \omega^\star_a(x)\>\delta_{ab}\> \delta\Gamma(x,y)\>
\omega_b(y) {\rm d}^dx{\rm d}^dy 
\label{dSgh}
\EE
together with its counterterm $\delta\Gamma$
\BE
\delta \Gamma(x,y)=
\left[{{\cal G}_0^{-1}}(x,y)-{{\cal G}_M^{-1}}(x,y)\right]
\label{dG}
\EE
where  ${{\cal G}_M}$ would be a massive ghost propagator
\BE
{{\cal G}_M}=\frac{1}{p^2-M^2}.
\label{GM}
\EE
One could wonder if the inclusion of a mass parameter in the trial ghost propagator could shift the pole of the
ghost at one-loop, yielding a phenomenological mass which would be at odds with the lattice data for the dressed ghost propagator.
However, in the massive expansion of the propagators\cite{ptqcd2,analyt} the
counterterm cancels the shift at tree level and any real mass term can only arise from loops. That is the reason why no
mass would arise for the photon in QED by the same method. It can be easily shown\cite{genself} that the ghost self energy is
of order ${\cal O}(p^2)$ and vanishes when the external momentum $p\to 0$, so that the dressed ghost propagator still has
a pole at $p^2=0$. That is an other way to see that the gluon mass arises from gluon loops in the expansion and is not
a mere shift by a mass parameter.

The case of a finite ghost trial-mass $M>0$ has been explored in Ref.\cite{comitini} and found to be sub-optimal
when compared with the standard choice of a massless ghost. Then, we will assume $M=0$ in the present variational study.
It must be mentioned that, if the ghost mass $M$ were
regarded as an independent variational parameter, then its stationary point would be at $M=0$ because there are no 
ghost-gluon vertices in the first order effective potential. Actually, the ghost contribution would be maximal at that
stationary point, because of the wrong sign of ghost statistics. However, as discussed in the next section, in the more general
context of the finite temperature formalism, a maximal ghost energy minimizes the eventual
weakening of Jensen-Feynman inequality that might occur in non-Abelian theories. 
While that weakening cannot be avoided entirely, we will suggest a rigorous 
way to control the error on the variational bound. Let us take aside the problem for a while and assume that the GEP can be trusted
as a variational method.

Since we have not changed the total action at all, we know that the sum of all graphs contributing to the longitudinal 
gluon polarization must give zero, because of gauge invariance. Thus, the exact longitudinal part of the gluon propagator 
must be equal to the free longitudinal propagator $\Delta^L_0(p)=\xi/(-p^2)$. While, in principle, $m_L$ could be used as a variational parameter, 
we expect that the best result is achieved if the trial $\Delta^L_m$ is taken to be equal to the exact $\Delta^L_0$ by 
setting $m_L=0$ in Eq.(\ref{Deltam}). 

Having set $M=m_L=0$, the variational ansatz becomes the same that was used in 
the massive expansion of Refs.\cite{ptqcd2,analyt,scaling} where no ghost and longitudinal masses
were inserted.
Only the pole of the transverse free-particle propagator is shifted
and compensated by inserting a transverse counterterm 
\BE
\delta\Gamma^{\mu\nu}(p)=-m^2\> t^{\mu\nu}(p)
\label{dG1}
\EE
among the vertices of the interaction, while the gauge-dependent longitudinal part of the gluon propagator is
left unchanged and equal to the exact result.
As shown in Ref.\cite{scaling}, that massive expansion is in very good agreement
with the data of lattice simulations. Moreover, that choice of counterterms has the merit of
providing a fully gauge invariant GEP at $T=0$, as shown below.

The calculation of the GEP follows the same steps as for the scalar theory. The GEP is obtained as
the first-order effective potential in the covariant formalism, including the counterterms among the interaction
vertices and in the limit of a vanishing background field, i.e. assuming that $\langle A_{a\mu}\rangle=0$ 
since gauge symmetry is not broken in the vacuum.
The effective action reads
\BE
e^{i\Gamma(a)}=\int_{1PI} {\cal D}_{A,\omega}\> e^{i S_0(a+A,\omega)+i S_{int}(a+A,\omega)} 
\EE
and the effective potential follows as $V=-\Gamma(0)/{\cal V}_4$ and is the sum of all connected 1PI vacuum graphs.
The first order graphs contributing to the GEP are shown in the second row of Fig.~\ref{fig1}.

The zeroth order gluon and ghost loops in Fig.~\ref{fig1} give
\BE
V_0=\frac{i}{2{\cal V}_4} \log {\rm Det} \Delta_{m}^{\mu\nu}- \frac{i}{{\cal V}_4} \log {\rm Det} {\cal G}_{0}.
\EE
The determinant of $\Delta_m^{\mu\nu}$ can be split as the product of determinants in the
orthogonal Lorentz subspaces, ${\rm Det} \Delta_m^{\mu\nu}={\rm Det}[\Delta_m^T\> t^{\mu\nu}] 
{\rm Det}[\Delta_0^L \ell^{\mu\nu}]$,
yielding
\BE
V_0=\frac{i(d-1)}{2{\cal V}_4}\Tr \log \Delta_m^T+
\frac{i}{2{\cal V}_4}\Tr \log \Delta_0^L
-\frac{i}{{\cal V}_4}\Tr \log {\cal G}_0.
\label{V0YM1}
\EE
where $d=4$ in a four dimensional space-time. 
 
The constant gauge dependent (infinite) term  $\Tr \log\xi$
is canceled by an equal factor in the normalization of the Faddeev-Popov functional,
so that using $\Delta_0^L/\xi=-{\cal G}_0$,
one-half of the ghost cancels the longitudinal term yielding
\BE
V_0(m)=N_A \left[(d-1) K(m)- K(0)\right]
\label{V0YM0}
\EE
where $N_A=N^2-1$. 

The crossed one-loop graphs in Fig.~\ref{fig1} are obtained by one insertion of the counterterms.
Since there are no ghost and longitudinal counterterms, there is only one crossed loop for the transverse
gluon. The identity Eq.(\ref{ide})  changes its sign for $\Delta_m^T$ and
inserting the counterterm of Eq.(\ref{dG1})
the sum of all one-loop graphs (zeroth and first order) can be written  as
\BE
V_{1L}(m)=\left( 1-m^2\frac{\partial}{\partial m^2}\right) V_0(m)
\EE
which reads
\BE
\frac{V_{1L}(m)}{N_A}=(d-1) \left[K(m)-\frac{1}{2} m^2 J(m)\right]- K(0).
\label{V1LYM10}
\EE
The functions $K(m)$ and $J(m)$ were defined in Eq.(\ref{I}) and their explicit regularized expression were
given in Eq.(\ref{IDR}). The formal result of Eq.(\ref{V1LYM10}) is gauge invariant and
also valid at finite temperature, since Eq.(\ref{ide}) still holds when the integrals $K$, $J$ acquire a thermal
part. 

The first-order effective potential also includes the two-loop gluon graph in Fig.~\ref{fig1}. 
For $m_L=0$ each loop of the longitudinal propagator contributes a factor $\xi J(0)$ which is zero by
dimensional regularization, so that the two-loop term is also gauge invariant at $T=0$. 
The same identical expression would be obtained in Landau gauge ($\xi=0$) if $m_L>0$. 
The calculation is formally different in the finite temperature formalism and will be studied in the next section. 
Here, we examine  the vacuum part that contributes to the GEP at $T=0$ and is relevant for discussing the issue of mass generation.
Inserting the seagull one-loop graph\cite{genself}
\BE
\Pi_{1L}=-\frac{(d-1)^2Ng^2}{d} J(m)
\label{Psea}
\EE
the two-loop term reads
\BE
V_{2L}(m)=\frac{N_ANg^2(d-1)^3}{4d} \left[J(m)\right]^2.
\label{V2LYM1}
\EE
Setting $d=4$ and adding the one-loop term of Eq.(\ref{V1LYM10}), in terms of the new effective coupling $\alpha$
\BE
\alpha=\frac{9N g^2}{32\pi^2}=\frac{9N}{8\pi}\alpha_s, \qquad \alpha_s=\frac{g^2}{4\pi}
\EE
a gauge invariant GEP is found that can be written as
\BE
\frac{V_{G}(m)}{3 N_A}=K(m)-\frac{m^2}{2} J(m)
+ 2\pi^2\alpha\left[ J(m)\right]^2
\label{GEPYM1}
\EE
having dropped the constant $K(0)$ which is zero at $T=0$. 
That is the same identical result obtained in Eq.(\ref{GEP}) for the scalar theory, provided
that the effective coupling $\alpha$ is replaced by $\lambda/(16\pi^2)$.
Thus, using the same dimensional regularization scheme of Section III and keeping $d>4$,
the renormalized GEP of Eq.(\ref{GEPsc}) is recovered in units of the optimal gluon-mass parameter $m_0$.
Inserting the correct normalization factor, the GEP reads
\BE
\frac{V_{G}(m)}{3 N_A}
=\frac{m^4}{128\pi^2} \left[\alpha \left(\log \frac{m^2}{m_0^2}\right)^2+2\log\frac{m^2}{m_0^2} -1\right]
\label{GEPym}
\EE
and was shown in Fig.~\ref{fig4}.
That figure shows the existence of two competing stationary points for the vacuum: an unstable
stationary point at $m=0$ and a stable minimum at $m=m_0$.

The existence of a stable massive vacuum is a remarkable non-perturbative prediction of the present
variational method and can be regarded as an argument for mass generation in pure Yang-Mills theory.
We are tempted to identify the unstable stationary point at $m=0$ with the massless scaling solution of 
Schwinger-Dyson equations. That solution  is not found in lattice simulations. 

In the next section, we will show that the two stationary points acquire a very different behavior at finite temperature.
The massless vacuum at $m=0$ develops a thermal mass that increases with temperature like for a standard massless boson, 
while the minimum at $m=m_0$ shows a decrease of the mass until a weak first order transition occurs before the merging
of the minima.

As shown in Fig.~\ref{fig4}, when written in physical units of $m_0$, 
the renormalized GEP is not very sensitive to the actual value
of the strong coupling $\alpha_s$, especially at the stationary points that might be identified as physical configurations.
Thus everything seems to be settled by the physical scale $m_0$, while the coupling $\alpha_s$ must be regarded
as a bare coupling at the scale $\Lambda_\epsilon$ according to our renormalization scheme discussed in Section \ref{sec3}.
Its actual value should be almost irrelevant and will be fixed by the principle of minimal sensitivity\cite{minimal}
as the stationary point of the critical temperature.

Since there is no scale in the original Lagrangian, the actual value of the mass $m_0$ cannot be predicted by the
theory and must come from the phenomenology. The massive expansion of Refs.\cite{ptqcd2,analyt} arises as the natural 
expansion around the best trial massive vacuum at $m=m_0$. By that expansion, at one loop, 
the gluon propagator was found in perfect
agreement with the data of lattice simulations\cite{scaling} in the Landau gauge.
The inverse dressing function, which is basically
given by the gluon self-energy, is determined without any free parameter and is not monotone,
with a pronounced minimum that allows us to fix the energy scale with good accuracy.
As shown in Fig.~3, the one-loop analytical expression for the running coupling reproduces the lattice
data very well. 
Sharing the same units of the lattice data in the Landau gauge, the scale
$m_0=0.73$ GeV is extracted for $N=3\>$ \cite{ptqcd2,scaling}. We will use that scale in the next sections.

\section{The GEP at finite temperature and deconfinement}\label{sec5}

At finite temperature, supposing that Jensen-Feynman inequality Eq.(\ref{JF}) holds, 
the first-order free energy  is 
bounded below by the exact free energy  ${\cal F}(T)$ 
that can be expressed as
\BE
e^{-\beta\left[{\cal V}_3\>{\cal F}(T)\right]} ={\cal Z}=\int {\cal D}_{A,\omega}\> e^{(S_0+S_{int})}
\label{Fex}
\EE
where the thermal action is the integral over imaginary time defined in Eq.(\ref{ThS}).
If we split the action as in the previous section, inserting the mass term Eq.(\ref{dS}) in the free part 
and the counterterm Eq.(\ref{dG1}) among the vertices, the free energy in Eq.(\ref{Fex}) is expanded by the
same formal massive expansion as before. The first-order approximation ${\cal F}_1(T,m)$ depends on the mass parameter $m$
and is given by the same graphs in the second row of Fig.~\ref{fig1}. When optimized it gives the GEP, while the optimal value
of $m$ that minimizes ${\cal F}_1(T,m)$ provides the best trial mass parameter $m(T)$ at finite temperature,
so that $m(0)=m_0$.

In non-Abelian theories, the GEP might be bounded below by an approximate free energy rather than the exact free energy.
Actually, the existence of ghosts in the covariant formalism and the appearance of states with negative norm 
in the Hamiltonian formalism might limit the use of Jensen-Feynman inequality Eq.(\ref{JF}) and 
Bogolubov's inequality Eq.(\ref{BI}), respectively, unless we have some physical evidence about the
safe cancellation of the unphysical degrees of freedom in the averages. However, we can show that a weaker form of 
Jensen-Feynman inequality still holds for the GEP.

The partition function in Eq.(\ref{Fex}) can be written as
\BE
{\cal Z}=\int {\cal D}_{A,\omega}\> e^{\displaystyle{S^{\prime}}} {\rm Det}{\cal M}_{FP}(A)
\label{Fex2}
\EE 
where ${\cal M}_{FP}(A)$ is the Faddev-Popov matrix, which is linear in the field $A^\mu_a$, and $S^\prime$ 
is the original total action without any ghost term, obtained by setting $\omega_a=0$ in the 
sum $S_0+S_{int}$. We can also define zeroth order free energy ${\cal F}^{\>\prime}_0$ and partition function
${\cal Z}^\prime_0$ without ghost terms as
\BE
e^{-\beta\left[{\cal V}_3\>{\cal F}^{\>\prime}_0\right]} ={\cal Z}^\prime_0=\int {\cal D}_{A}\> e^{S^\prime_0}
\label{F0}
\EE
where $S^\prime_0$ is the quadratic part of $S^\prime$, including the gluon-mass term. 
The exact free energy ${\cal F}_{exact}$ follows as
\BE
{\cal F}_{exact}=
{\cal F}^{\>\prime}_0-T\log\left\langle e^{ {\displaystyle S^\prime}\!_{\! int} }\>{\rm Det}{\cal M}_{FP}(A)\right\rangle_0
\label{Fex3}
\EE
where $S^\prime_{int}=S^\prime-S^\prime_0$ and the average over $A_a^\mu$ is defined according to
\BE
\langle\dots\rangle_0=\frac{1}{{\cal Z}^\prime_0}\int {\cal D}_{A}\> e^{S^\prime_0}(\dots).
\EE
In Eq.(\ref{Fex3}), we can use Jensen inequality  in the pure bosonic average of the convex exponential function and write
\BE
{\cal F}_{exact}\leq {\cal F}^{\>\prime}_1+{\cal F}^{gh}
\label{vartrue}
\EE
where
\BE
{\cal F}^{\>\prime}_1={\cal F}^{\>\prime}_0-T\left\langle  S^\prime_{int}\right\rangle_0
\EE
is the sum of all first-order gluon graphs in the second row of Fig.~\ref{fig1} and gives the gluon contribution to
the first-order free energy, while  ${\cal F}^{gh}$ is a ghost free-energy given by
\BE
{\cal F}^{gh}=-T\left\langle \log {\rm Det}{\cal M}_{FP}(A)  \right\rangle_0
\label{Fgh}
\EE
which is different from the sum of all first-order ghost graphs ${\cal F}^{gh}_1$ contributing to the GEP in Fig.~\ref{fig1}.
If the ghost term ${\cal F}^{gh}$ were known exactly, then its sum with the gluon first-order term  ${\cal F}^{\>\prime}_1$ 
would provide through Eq.(\ref{vartrue}) a pure variational approximation, bounded below by the exact free energy.

We can loop expand ${\cal F}^{gh}$ by inserting the explicit form of the matrix ${\cal M}_{FP}$. In any linear covariant
gauge
\BE
{\cal M}_{FP}(A)={\cal G}^{-1}_M+\delta {\cal M}(A)
\EE
where the massive ghost propagator was defined in Eq.(\ref{GM}) and takes account of a generic shift of the pole, while
$\delta {\cal M}(A)$ is the sum of the ghost vertex of ${\cal L}_{gh}$ in Eq.(\ref{LintYM}) (proportional to the gauge
field $A^\mu_a$) and the ghost counterterm $\delta \Gamma$ of Eq.(\ref{dG}). Expanding the log we obtain
\begin{align}
\beta {\cal F}^{gh}&=\Tr \log {\cal G}_M-\Tr\left({\cal G}_M\delta\Gamma\right)\nn\\
&+\frac{1}{2}\langle \Tr\left[{\cal G}_M\delta {\cal M}(A){\cal G}_M\delta {\cal M}(A)\right]\rangle_0+\dots
\end{align}
which is a sum of vacuum ghost graphs with insertions of the standard vertices. The first two terms of the expansion
are just the first-order ghost graphs in Fig.~\ref{fig1} and give the ghost term ${\cal F}^{gh}_1$ contributing to the GEP.
The third term is the two-loop graph
\BE
{\cal F}^{gh}_{2L}\sim\alpha\int {\cal G}_M\Delta_m{\cal G}_M
\label{2LFgh}
\EE
which might be added to the first-order terms for improving the approximation, as discussed by previous work in the
Lagrangian and Hamiltonian formalism\cite{HT,reinhardt05}. We observe that, while the bound in Eq.(\ref{vartrue}) is
exact, any arbitrary truncation of the expansion would invalidate it. Thus, there is no way to tell if adding the two-loop
term would give a better result compared with the simple GEP where only the first-order terms are retained.
Denoting by $\delta{\cal F}$ the difference between the exact ghost term and the first-order terms retained in the GEP
\BE
\delta{\cal F}={\cal F}^{gh}-{\cal F}^{gh}_1
\label{delta}
\EE 
We can write the exact bound in Eq.(\ref{vartrue}) as
\BE
{\cal F}_{GEP}={\cal F}^{\prime}_1+{\cal F}^{gh}_1\geq  {\cal F}_{exact}-\delta {\cal F}.
\label{varfalse}
\EE
The GEP might actually fall below the exact free energy, but we can minimize the problem by {\it maximizing}
the ghost term ${\cal F}^{gh}_1$ in the GEP, as suggested by Eq.(\ref{delta}).
In fact, it can be easily shown that $\delta {\cal F}\geq 0$ and ${\cal F}^{gh}_1$ is bounded above by the
exact ghost term ${\cal F}^{gh}$. By use of Jensen inequality in the average of the log in Eq.(\ref{Fgh})
\begin{align}
{\cal F}^{gh}&\geq -T \left[\Tr \log \left\langle{\cal M}_{FP}(A)  \right\rangle_0\right]\nn\\
&=T\left[\Tr \log {\cal G}_0\right]={\cal F}^{gh}_1\bigg\vert_{M=0}
\end{align}
and since ${\cal F}^{gh}_1$ is maximal at its stationary point $M=0$, that point is also the safest choice that
maximizes the ghost term without reaching the exact value ${\cal F}^{gh}$. 
Having shown that $\delta {\cal F}$ is positive, we could estimate its value by an explicit evaluation of the two-loop
term in Eq.(\ref{2LFgh}) in order to keep the approximation under control. We must mention that the GEP might be closer
to the exact free energy than expected by the mathematical bound of Eq.(\ref{varfalse}) since $\delta {\cal F}$ is just
the maximal error that we have been able to establish in the worst case. In fact, by a comparison with the data
of lattice simulations, we will show that at finite temperature the GEP does very well, better than expected by the present
analysis.

At finite temperature, the explicit calculation of the GEP follows by the graphs of Fig.~\ref{fig1}.
The sum of one-loop graphs is still given by Eqs.(\ref{V1LYM10}) where the integrals $K$, $J$ in Eq.(\ref{KJ}) now
include a sum over discrete frequencies and their explicit expressions in Eq.(\ref{I}) are replaced by
\begin{align}
K(T,m)&=\frac{1}{2}\sip \log({\bf p}^2+\omega_n^2+m^2)\nn\\
J(T,m)&=\sip \frac{1}{{\bf p}^2+\omega_n^2+m^2}
\label{Ith}
\end{align}
having used in Eq.(\ref{KJ}) the massive free propagator
\BE
\Delta_m(\omega_n,{\bf p})=\frac{1}{{\bf p}^2+\omega_n^2+m^2}
\label{Delta}
\EE
in the Euclidean space where $p^\mu=(\omega_n, {\bf p})$ and $\omega_n=2\pi n T$.
In the limit $T\to 0$ the vacuum integrals in Eqs.(\ref{I}) are recovered as
$J(m)=J(0,m)$ and $K(m)=K(0,m)$. We denote them by $J_V(m)$ and $K_V(m)$, respectively.
They contain the diverging part of the integrals and can be regularized as discussed in the
previous sections. Their explicit expression is given by Eqs.(\ref{IDR}).
The thermal parts are finite but depend on $T$. We denote them by $J_T(T,m)$ and $K_T(T,m)$, respectively.
Omitting the arguments for brevity, they can be written by an explicit calculation as
\begin{align}
K_{T}&=K-K_V=-\frac{1}{6\pi^2}\int_0^\infty \frac{ n(\epsilon_{k,m})}{\epsilon_{k,m}} k^4 {\rm d}k\nn\\
J_{T}&=J-J_V=\frac{1}{2\pi^2}\int_0^\infty \frac{ n(\epsilon_{k,m})}{\epsilon_{k,m}} k^2 {\rm d}k
\end{align}
where $\epsilon_{k,m}=\sqrt{k^2+m^2}$ and $n(\epsilon)=\left[\exp(\beta\epsilon)-1\right]^{-1}$ is the Bose distribution.

The first-order free energy ${\cal F}_1(T,m)$ can be written as the sum of one-loop and two-loop
terms
\BE
{\cal F}_1(T,m)={\cal F}_{1L}(T,m)+{\cal F}_{2L}(T,m).
\label{F1}
\EE
The sum of one-loop graphs is obtained by just setting $d=4$ in Eq.(\ref{V1LYM10})
\begin{align}
{\cal F}_{1L}(T,m)&=3N_A\left[K(T,m)-\frac{1}{2} m^2 J(T,m)\right]\nn\\
&-N_A\> K(T,0).
\label{F1L}
\end{align}
The second term ${\cal F}_{2L}(T,m)$ is the two-loop graph in the second row of Fig.~\ref{fig1}. 
Because of the breaking of Lorentz
invariance at finite $T$, its expression gets formally different than the vacuum term in Eq.(\ref{V2LYM1}) and
also becomes gauge dependent. In order to make contact with previous analytical and numerical work in the Landau gauge
we set $\xi=0$, which is the most common choice for the study of the correlators, so that the scale $m_0=0.73$ GeV will
be used. In fact, that scale was extracted by matching the predictions of the massive expansion with the data of numerical simulations
in the Landau gauge\cite{ptqcd2,scaling}. 
Assessing the whole gauge dependence of the GEP at finite temperature is not an easy task, as the scale $m_0$ should be also changed by matching the gauge-dependent correlators in a different gauge.

Following the same steps of the previous sections, in the Landau gauge, the seagull graph of the gluon self energy can be 
written as\cite{genself}
\BE
\Pi_{ab}^{\mu\nu}=-\delta_{ab} Ng^2\sip\left[2\delta^{\mu\nu}\Delta_m+p^\mu p^\nu\Delta_0\Delta_m\right]
\EE
where $\Delta_m=\Delta_m(p)$ is the Euclidean propagator in Eq.(\ref{Delta}). 
Integrating the single terms, it can be written as
\BE
\Pi_{ab}^{\mu\nu}=-\delta_{ab} Ng^2\left[2\delta^{\mu\nu}J+I^{\mu\nu}\right]
\label{Psea2}
\EE
where
\BE
I^{\mu\nu}=\sip p^\mu p^\nu \Delta_m(p)\Delta_0(p).
\EE
The trace of $I^{\mu\nu}$ is $I^{\mu\mu}=J$, so that at $T=0$, by Lorentz invariance, 
the self energy of Eq.(\ref{Psea}) is recovered for $d=4$. At finite temperature, $I^{\mu\nu}$ is still diagonal
but $I^{00}\not= I^{ii}$. By rotational invariance, using the trace again, we can write
\BE
I^{11}=I^{22}=I^{33}=\frac{1}{3}\left(J-I^{00}\right)
\EE
which holds separately for the thermal and vacuum parts.
While the vacuum part is just $I_V^{00}=I_V^{ii}=J_V/4$, the thermal part can be obtained by an explicit
integration as
\BE
I^{00}_T=\frac{1}{m^2}\left(h_m-h_0\right)
\EE
where $h_m$ is the integral
\BE
h_m=\frac{1}{2\pi^2}\int_0^\infty  \epsilon_{k,m}\> n(\epsilon_{k,m}) k^2 {\rm d}k
\EE
that can be evaluated exactly for $m=0$ yielding 
\BE
h_0=-3K_T(T,0)=\frac{\pi^2 T^4}{30}.
\label{hK0}
\EE

Closing the second loop with the transverse  gluon propagator ($\xi=0$) and inserting the symmetry factor $1/4$
\BE
{\cal F}_{2L}=-\frac{1}{4}\Pi_{ab}^{\mu\nu}\>\sip \Delta_m(p) t_{\mu\nu}(p)\delta_{ab}.
\EE
Then, using Eq.(\ref{Psea2}), the two-loop term reads
\BE
{\cal F}_{2L}=\frac{N_A Ng^2}{4} \left( 7J^2-I^{\mu\nu} I^{\mu\nu}\right)
\label{F2L}
\EE
and its inclusion in Eq.(\ref{F1}) together with Eq.(\ref{F1L}) gives the first-order free energy in closed form. When optimized, it provides the GEP
at finite temperature. With some abuse of language we can denote the first-order free energy by ${\cal F}_G (T,m)$ and call 
it the GEP.

It is useful to separate the thermal and vacuum parts of the GEP. If we do that and use the explicit
regularized expressions Eqs.(\ref{IDR}) for the vacuum parts $J_V$, $K_V$, the total first-order free energy of Eqs.(\ref{F1}),(\ref{F1L}),
(\ref{F2L}) can be easily shown to become
\BE
{\cal F}_G (T,m)={\cal F}_G (0,m)+\Delta {\cal F}_G (T,m)
\label{GEPT}
\EE
where the vacuum part ${\cal F}_G (0,m)=V_G(m)$ is just the GEP at $T=0$, given by Eq.(\ref{GEPym}) when expressed in
terms of $m_0$. The thermal part $\Delta {\cal F}_G (T,m)$ vanishes at $T=0$ and can be written as
\begin{align}
\frac{\Delta {\cal F}_G (T,m)}{3N_A}&=K_T+\frac{\pi^2}{270} T^4
+\frac{\alpha \>m^2}{4}\> J_T \>\log\frac{m^2}{m^2_0}\nn\\ 
&+ 2\pi^2\alpha \left[ J_T^2- \left(\frac{2}{3}\right)^4\left(\frac{J_T}{4}-I_T^{00}\right)^2\right].
\label{GEPT2}
\end{align}

\begin{figure}[t] 
\centering
\hspace*{-1cm}\includegraphics[width=0.4\textwidth,angle=-90]{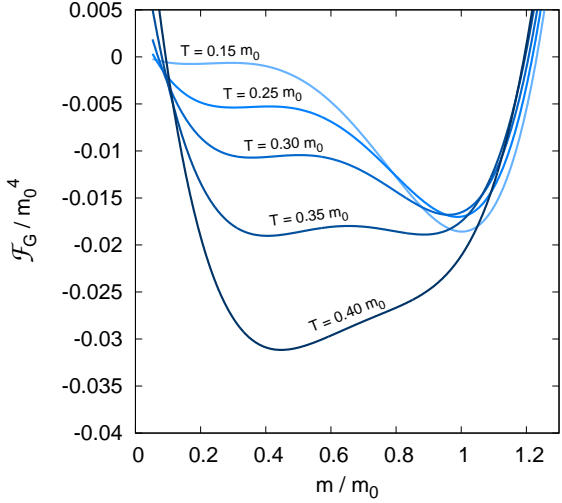}
\caption{The renormalized GEP of Eqs.(\ref{GEPT}),(\ref{GEPT2}) is shown in units of $m_0$ for $\alpha_s=0.9$
and different values of the temperature.}  
\label{fig5}
\end{figure}

\begin{figure}[t] 
\centering
\hspace*{-1cm}\includegraphics[width=0.4\textwidth,angle=-90]{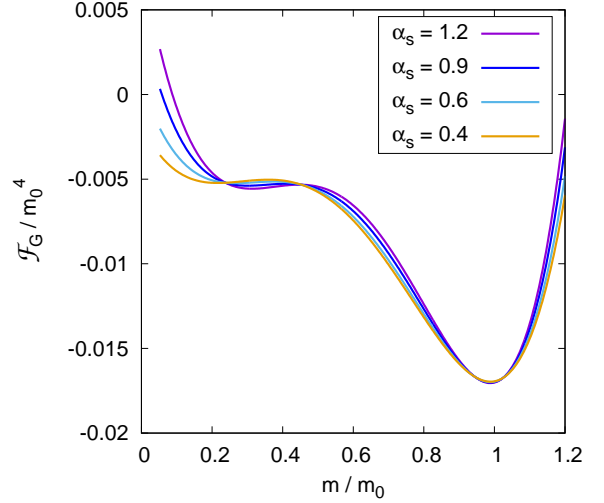}
\caption{The renormalized GEP of Eqs.(\ref{GEPT}),(\ref{GEPT2}) is shown in units of $m_0$ for $T/m_0=0.25$ and
different values of the strong coupling $\alpha_s$.}  
\label{fig6}
\end{figure}

The GEP is shown in Fig.~\ref{fig5} for different values of the temperature and  in Fig.~\ref{fig6}
for several values of the coupling $\alpha_s$.  As already discussed in the previous sections, the GEP is not
very sensitive to the coupling, especially in the physical ranges around the minima and for $T< 2T_c\approx 0.5$ GeV.

While the physical value of the GEP was not sensitive at all to a change of $\alpha_s$ at $T=0$, 
other observables, at finite temperature, might depend on $\alpha_s$ because the variational method is not an exact calculation.
In lattice simulations, the bare coupling and the cutoff are finite, since the lattice spacing cannot be set to zero.
However, a stationary regime is reached where the physical predictions seem to be not sensitive to the actual value of the
bare coupling. In the present calculation, because of the approximations, we fail to reach an exactly stationary regime for
all the thermal observables. Albeit small, a residual sensitivity to the bare $\alpha_s$ is found, posing the problem of the
choice of the coupling. We argue that, for any finite value of coupling and cutoff, the outcome of the variational calculation
is more reliable and closer to the lattice data if the physical observables are less sensitive to the arbitrary value
of the bare coupling. 
Thus, the best agreement with the data of lattice simulations is expected in the range $0.6<\alpha_s<1.2$ where a
real plateau is observed, rather than in the limit $\alpha_s\to 0$ where a slightly larger sensitivity is found.
For that reason, even if $\alpha_s$ should be sent to zero in the limit $\epsilon\to 0$,
we prefer to keep $\alpha_s$ fixed at the optimal value $\alpha_s=0.9$ in the following discussion and in the comparison
with the lattice data. We checked that any other choice does not introduce important changes in the results.

\begin{figure}[h] 
\centering
\includegraphics[width=0.35\textwidth,angle=-90]{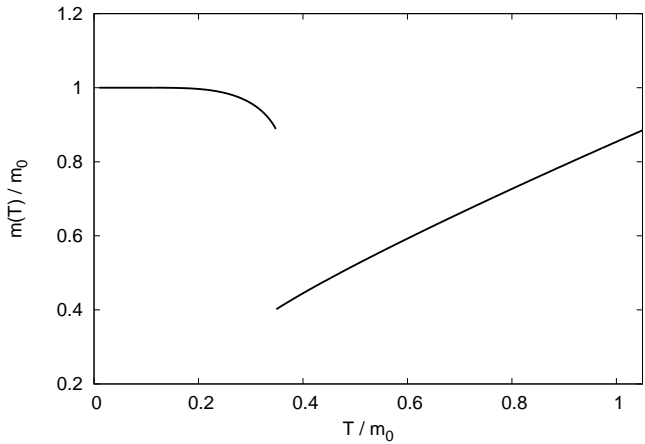}
\caption{The optimal mass parameter $m(T)$ which minimizes the GEP is shown as a function of temperature
for $\alpha_s=0.9$.}
\label{fig7}
\end{figure}

\begin{figure}[h] 
\centering
\hspace*{-1cm}\includegraphics[width=0.4\textwidth,angle=-90]{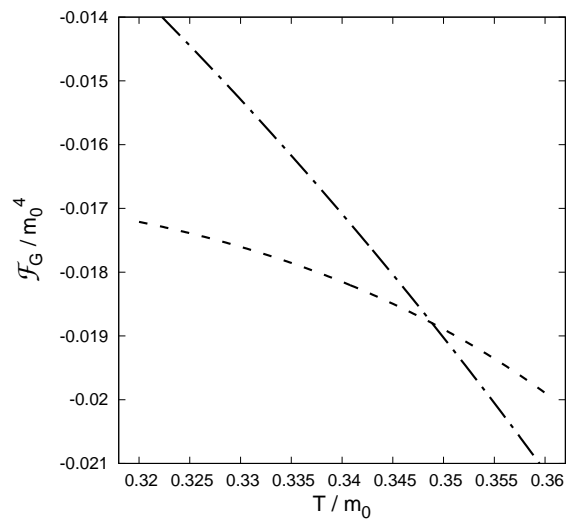}
\caption{The Free energy at the minima of the GEP across the transition for $\alpha_s=0.9$. 
Below the transition point, the upper curve (dot-dashed) is the GEP at the unstable thermal mass while the lower
curve (dashed) is the GEP at the stable dynamical mass. The order reverses above the transition point.}
\label{fig8}
\end{figure}

At finite temperature, we observe that the minima of the GEP have a very different behavior. The absolute minimum at
$m=m_0$ is almost frozen when $T\ll m_0$, as expected for a massive confined gluon. When the temperature increases the
minimum moves backwards, so that the optimal mass parameter $m(T)$ is a decreasing function of the temperature, in fair
agreement with the decrease of mass that is observed on the lattice below $T_c$\cite{silva}.
The unstable minimum, at $m=0$ in Fig.~\ref{fig4}, moves forward when $T>0$ and its mass value increases almost 
linearly like the thermal mass of a massless boson. 
It gets deeper with increasing temperature. Thus the GEP seems to show the competition between
a confined boson with a dynamical mass and a free boson with a thermal mass.
As shown in Fig.~\ref{fig5}, at a critical temperature $T_c\approx 0.35 m_0$ the minima reach the same free energy
before they can merge, so that a weak first-order  phase transition is predicted with a discontinuous drop of the
optimal mass parameter $m(T)$ that is displayed in Fig.~\ref{fig7}. 
The free energy at the minima is shown in Fig.~\ref{fig8}
across the transition. Below the transition point, the upper curve is the GEP at the unstable thermal mass, 
while the lower curve is the GEP at the stable dynamical mass. 
Above the transition point they reverse. At any temperature, the physical 
free energy is the lower curve ${\cal F}_G( T, m(T))$.

\begin{figure}[h] 
\centering
\hspace*{-0.5cm}\includegraphics[width=0.4\textwidth,angle=-90]{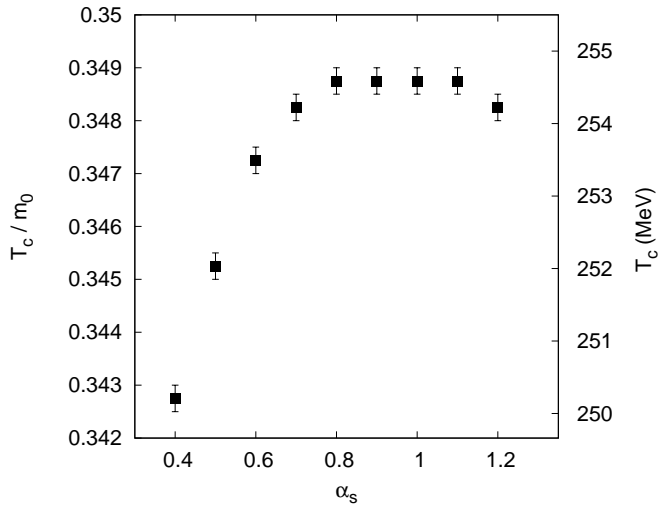}
\caption{The critical temperature is shown at a very enlarged scale, as a function
of the coupling $\alpha_s$. The minor effect of its change is less than $\pm 1\%$.
Error bars are the numerical error in the calculation. The right hand scale is obtained
by taking $m_0=0.73$ GeV.} 
\label{fig9}
\end{figure}

The slight effect of a change of $\alpha_s$ on the critical temperature is less than $\pm 1\%$ 
in Fig.~\ref{fig9}, where it is shown at a very enlarged scale. Apart the effect of the scale, the critical temperature
is basically unchanged for a large range of $\alpha_s$, including the phenomenological interval $0.4<\alpha_s<1.2$ 
which would be ranged by a running coupling in the IR. The plateau has a stationary point at $\alpha_s\approx 0.9$
where $T_c=0.349 \>m_0$. We take that as the best prediction of the GEP according to the 
principle of minimal sensitivity\cite{minimal}.

Using the scale $m_0=0.73$ GeV that arises for $N=3$ from the massive expansion at
one-loop\cite{ptqcd,ptqcd2,analyt,journey,damp,scaling}, we predict $T_c=255$ MeV, 
which is very close to the value $T_c=270$ MeV that is found on the lattice\cite{silva}. 

It is important to mention that if the bare coupling were sent to zero in the limit $\epsilon\to 0$,
the resulting qualitative picture would remain basically unchanged.
In the limit $\alpha_{s}\to 0$, the deconfinement transition still takes place, 
is weakly first order and with a critical temperature $T_{c}\approx0.32\,m_{0}$ not too far from that found on the plateau. 
The only relevant difference is in the behavior of the unstable minimum, 
whose position does not change with the temperature and 
remains fixed at $m=0$ for every value of $T$, even if it gets deeper and eventually becomes the stable minimum above $T_c$. 
Thus, in the limit $\alpha_{s}\to 0$, the optimal mass parameter is $m\approx m_{0}$ for $T< T_{c}$, and $m=0$ for $T>T_{c}$.
In the same limit, the critical temperature can be estimated by observing that the gluon thermal term is
exponentially suppressed at $m\approx m_0$ and cancels the opposite ghost term, so that 
the minimum of ${\cal F}_G(T,m)$ is basically frozen at the vacuum value
${\cal F}_G(T,m)\approx V_G(m_0)=-3N_A m_0^4/(128 \pi^2)$ if $T\ll m_0$. 
On the other hand, 
setting $\alpha=0$ in Eq.(\ref{GEPT2}), the unstable minimum at $m=0$ is given by ${\cal F}_G(T,0)=-3N_A\pi^2 T^4/135$, 
so that a first order phase transition occurs at 
$T_c\approx\left(\frac{135}{128}\right)^{\frac{1}{4}}\frac{m_0}{\pi}=0.32\,m_0$ 
where the optimal mass parameter drops to zero.

The equation of state can be studied by introducing
pressure and entropy density according to
\begin{align}
p&=-\left[{\cal F}_G( T, m(T))-{\cal F}_G( 0, m_0)\right]\nn\\
s&=-\frac{\partial}{\partial T}{\cal F}_G( T, m(T)).
\label{EoS}
\end{align}

The reader might have noticed in Fig.~\ref{fig5} that below $T_c$ the minimum at $m=m_0$ moves slightly upwards.
That behaviour gives an unphysical negative entropy for a limited range of temperatures, as reported by other
massive approximations at one-loop\cite{tissierplb,tissier15} and by other variational methods\cite{quandt}. 
That minor shortcoming might be expected since
the contribution of the massless ghost is enhanced when $T\ll m$ compared to the massive gluon. 
The problem becomes more evident if we look at the ratio $p/ T^4$ in the limit $T\to 0$. That ratio
should be exponentially suppressed and dominated by the lightest glueball mass, in agreement with the data
of lattice simulations\cite{giusti,eos12,boyd96}. By inspection of Eq.(\ref{GEPT2}), we observe that while  
the thermal functions $K_T$, $J_T$, $I^{00}_T$ are exponentially suppressed, the second term on the right hand side
contributes with the fourth power of $T$, originating from the massless ghost loop in Eq.(\ref{hK0}) which, besides,
is taken with the opposite sign. When all other terms are suppressed, the ghost loop dominates the leading behavior
yielding a finite non-zero ratio in the limit $T\to 0$
\BE
\frac{p}{T^4}\> \to\> -\frac{N_A\pi^2}{90}
\label{pT0}
\EE
and a negative entropy in the same limit. That seems to be a shortcoming of the Landau gauge, since the
same identical finite values were found in Refs.\cite{tissiersu2,tissier15} in that gauge. The same authors
find smaller finite values and a positive entropy in the Landau-De Witt gauge by a two-loop calculation.
As discussed in Ref.\cite{quandt}, one would be tempted to cancel the unphysical term by hand, but that term 
gives an important contribution above the transition where it cancels unphysical gluon terms.

\begin{figure}[h] 
\centering
\hspace*{-1cm}\includegraphics[width=0.4\textwidth,angle=-90]{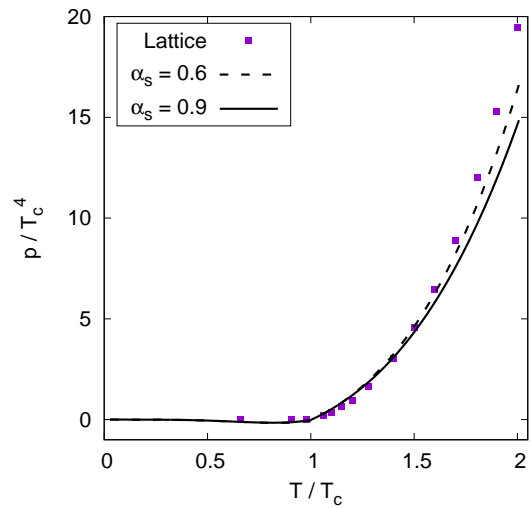}
\caption{Equation of state. The pressure is evaluated by Eq.(\ref{EoS}) and shown in units of $T_c$ 
for the optimal coupling $\alpha_s=0.9$ (solid line) and for $\alpha_s=0.6$ (broken line). 
The squares are the lattice data of Ref.\cite{giusti}.}
\label{fig10}
\end{figure}

On the other hand, the mismatch can only be observed below $T_c$ where the exact free-energy is almost constant and the pressure
is basically zero, so that even a very small (positive) deviation can give an increasing free-energy and a decreasing
pressure. Actually, the effect can be hardly seen in Fig.~\ref{fig10} where the pressure of Eq.(\ref{EoS})
is shown together with
the recent lattice data of Ref.\cite{giusti} which are consistent with previous existing data\cite{eos12,boyd96}. 
We observe that the figure is not a fit and that there are no free parameters in the calculation. Moreover, in units of
$T_c$ the pressure in Fig.~\ref{fig10} does not even depend on the energy scale $m_0$. Thus, it is
remarkable that the data points fall so close to the prediction of the calculation, at least for $T<2 T_c$.
As shown in the figure, the GEP  provides a pressure that seems to be bounded above by the data points, 
as expected if the GEP were bounded below by the exact free energy, suggesting that the error in the ghost
free-energy $\delta {\cal F}$ might be very small in Eq.(\ref{varfalse}).
For comparison, in Fig.~\ref{fig10} the pressure is also shown for a coupling $\alpha_s=0.6$, smaller than the optimal
value ${\alpha_s=0.9}$. While the predictions are not sensitive to the choice of the coupling at low temperature,
above $1.5\> T_c$ the pressure acquires a slight dependence on it and the agreement with the data
improves by decreasing $\alpha_s$.

\begin{figure}[h] 
\centering
\hspace*{-1cm}\includegraphics[width=0.4\textwidth,angle=-90]{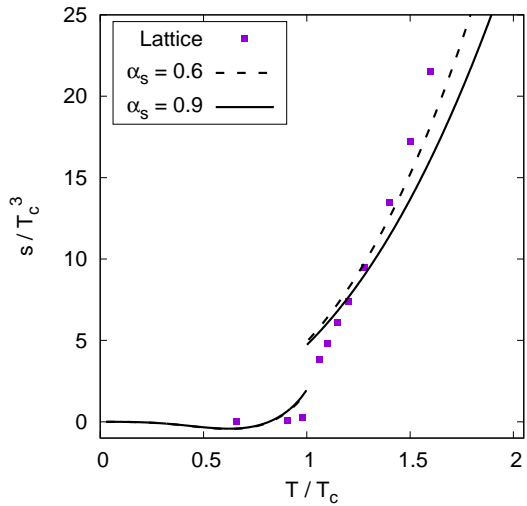}
\caption{Equation of state.The entropy density is evaluated by Eq.(\ref{EoS}) and shown in units of $T_c$
for the optimal coupling $\alpha_s=0.9$ (solid line) and for $\alpha_s=0.6$ (broken line). The squares are the
lattice data of Ref.\cite{giusti}.}
\label{fig11}
\end{figure}

The problem of a negative entropy becomes more evident in Fig.~\ref{fig11} where the entropy density of Eq.(\ref{EoS})
is shown together with the lattice data of Ref.\cite{giusti}. 
The small jump of the entropy density at $T=T_c$ is $\Delta s/T_c^3=2.7$ yielding a latent heat $\Delta H_0=2.7\> T_c^4$
which is larger than the values $1.3-1.5$ found in lattice simulations\cite{giusti,eos12,boyd96}.
However, we expect that the overall picture of dynamical mass generation, deconfinement transition and equation of state
might improve greatly by adding higher-order terms of the expansion in the free energy, 
as it is the case for the dressed propagator which gets on top of the lattice data when the one-loop 
terms are added to the zeroth-order massive propagator $\Delta_m=1/(p^2+m_0^2)$\cite{ptqcd2,analyt,scaling}.

\section{Discussion}\label{sec6}

The self-consistency gap equation of the GEP, Eq.(\ref{gap}) has attracted a lot of attention 
in the past\cite{papa08,papa10,cornwall82} 
as a basic physical tool for explaining the dynamical mass generation of Yang-Mills theories.
The main difficulty of handling the gap equation has always been the regularization of the diverging
integral $J(m)$ and its physical meaning. Here, we have shown that, by dimensional regularization in $d>4$,
the GEP provides a reasonable account of the general features of Yang-Mills theory.
The existence of a deep minimum at $m=m_0\not=0$ can be regarded as a variational argument for
dynamical mass generation in the original scale-less theory.

In order to enforce our confidence on the genuine physical nature of the minimum, we explored the model at finite temperature.
The emerging scenario for the equation of state and the deconfinement transition is in very good agreement
with the data of lattice simulations, leaving no doubt about the physical interpretation of the minima in the GEP.

Moreover, the method provides a perturbative tool for improving the results order by order. The expansion around the optimal
vacuum of the GEP turns out to be the massive expansion developed in Refs.\cite{ptqcd,ptqcd2,analyt} which provides accurate
and analytical expressions for the propagators at one-loop already. Once the non-perturbative effects are embedded in the
optimal variational mass, the residual interaction can be described by perturbation theory yielding a powerful 
analytical tool for QCD in the IR.

Thus, we argue that the present variational estimate of the thermodynamical potentials might be improved by inclusion of
higher order terms. 
Second order extensions of
the GEP have been discussed by several authors\cite{stancu2,gep2,HT,stancu}. In general, they do not retain the genuine 
variational property of the GEP but different optimization strategies have been proposed ranging from the principle of
minimal sensitivity\cite{minimal} to the method of minimal variance\cite{sigma2,varqed,varqcd,genself}. Explicit massive 
two-loop thermal graphs have been evaluated in Ref.\cite{tissier15}.
Here, we limited the calculation at the first order, just because we preferred to maintain the genuine variational nature of
the method unspoiled, as much as Jensen-Feynman inequality allows in presence of ghost fields.
Nevertheless, the pure GEP provides a remarkably good picture of the deconfinement transition.
From first principles, without any fit parameter, the simple first-order calculation predicts a weak first order transition
at $T_c\approx 250$ MeV for $N=3$, with a pressure which is very close to the data points of lattice simulations.
 We must mention that the method fails to predict a continuous transition for $N=2$. That could be the
consequence of a known issue for the GEP which usually predicts a weak first-order transition even when the transition
is second-order, e.g. for the scalar theory\cite{stevenson,stancu2}. In that case, a continuous transition is restored by inclusion
of second order terms\cite{stancu2}. Moreover, the GEP is known\cite{ibanez,stevensonN} 
to predict the correct $N\to\infty$ limit of $1/N$ expansions, so that its reliability increases when $N$ is large.

Finally, even if the present variational study is limited to the low temperature range $T<m_0$, 
where no resummation of hard thermal loops is required because of the finite mass in the loops,  the effects 
of a finite mass become negligible for large energies and $T\gg m_0$  and the standard results of perturbation theory 
would be recovered by the massive expansion in that limit.

\acknowledgments

We thank V. Branchina, M. Consoli and U. Reinosa for helpful discussions and feedback
on the manuscript.

\end{document}